\documentclass[twocolumn]{aastex631}
\usepackage{amssymb, amsmath, amsbsy, booktabs,subfigure}
\usepackage{CJK}
\usepackage{anyfontsize}

\newcommand{\lbol}{\ensuremath{L\mathrm{_{bol}}}}

\newcommand{\ledd}{\ensuremath{L\mathrm{_{Edd}}}}
\newcommand{\redd}{\ensuremath{L_\mathrm{bol}/L_{\mathrm{Edd}}}}

\newcommand{\lfive}{\ensuremath{\lambda L_{\lambda}(5100}~\AA)}

\newcommand{\msun}{\ensuremath{M_{\odot}}}

\newcommand{\ergs}{\ensuremath{\mathrm{erg~s^{-1}}}}
\newcommand{\kms}{\ensuremath{\mathrm{km~s^{-1}}}}
\newcommand{\mbh}{\ensuremath{M_\mathrm{BH}}}

\newcommand{\chisq}{\ensuremath{\chi^2}}

\newcommand{\ha}{{\rm H\ensuremath{\alpha}}}
\newcommand{\hb}{{\rm H\ensuremath{\beta}}}

\newcommand{\bha}{H\ensuremath{\alpha}}

\newcommand{\hii}{H\,{\footnotesize II}}
\newcommand{\nii}{[N\,{\footnotesize II}]}
\newcommand{\sii}{[S\,{\footnotesize II}]}
\newcommand{\oiii}{[O\,{\footnotesize III}]}

\newcommand{\feii}{{\rm Fe\,{\footnotesize II}}}

\newcommand{\mgii}{Mg\,{\footnotesize II}}

\newcommand{\oi}{[O\,{\footnotesize I}]}

\newcommand{\caii}{Ca\,{\footnotesize II}}

\newcommand{\sixpt}{\fontsize{6pt}{7.2pt}\selectfont}

\begin{document}
 
\title{A Sample of Active Galactic Nuclei with Intermediate-mass Black Holes Extended to $z \approx$ 0.6}

\author{Wen-Juan Liu \begin{CJK*}{UTF8}{gbsn} (刘文娟) \end{CJK*} }
\affiliation{Yunnan Observatories, Chinese Academy of Sciences, 
Kunming, Yunnan 650011, China; wjliu@ynao.ac.cn}

\author{Luis C. Ho}
\affiliation{Kavli Institute for Astronomy and Astrophysics, Peking University, 
Beijing 100871, China; }
\affiliation{Department of Astronomy, School of Physics, Peking University, 
Beijing 100871, China}

\author{Xiao-Bo Dong \begin{CJK*}{UTF8}{gbsn} (董小波) \end{CJK*}}
\affiliation{Yunnan Observatories, Chinese Academy of Sciences, 
Kunming, Yunnan 650011, China; wjliu@ynao.ac.cn}

\author{Su Yao \begin{CJK*}{UTF8}{gbsn} (姚苏) \end{CJK*}}
\affiliation{National Astronomical Observatories, Chinese Academy of Sciences,
Beijing 100101, China}

\author{Paulina Lira}
\affiliation{Departamento de Astronom\'ia, Universidad de Chile, Casilla 36-D, 
Santiago, Chile}

\author{Yicheng Guo}
\affiliation{Department of Physics and Astronomy, University of Missouri, 
Columbia, MO 65211, USA}



\begin{abstract}
We present a sample of 930 intermediate-mass black hole active galactic nuclei (AGNs) 
with BH masses of $\mbh\leqslant2\times10^{6}$ \msun,
uniformly selected from the 
Seventeenth Data Release of the Sloan Digital Sky Survey,
based on the detection of broad \ha\ or \hb\ emission lines.
Taking advantage of the wide wavelength coverage of BOSS/eBOSS spectroscopy, 
our sample extends the redshift coverage of low-$z$ IMBH AGNs to $z\leqslant0.57$,
significantly improving upon previous studies that were generally limited to $z\leqslant0.35$.
This sample encompasses BH masses from $10^{4.0}$ to $10^{6.3}$ \msun,
with Eddington ratios ranging from 0.01 to 1.9.
Among the $z>0.3$ subset, 24 sources exhibit detectable broad \mgii\,$\lambda\lambda2796,2803$ emission lines, 
including eight confirmed by independent DESI spectra.
A preliminary analysis reveals a marked decline in both the maximum accretion rate
(\redd) and broad \ha\ luminosity with decreasing redshift,
possibly reflecting a cosmic evolution of accretion activity at the low-mass end,
akin to the ``downsizing'' evolutionary trend seen in high-mass AGNs.

\end{abstract}



\section{Introduction}

Supermassive black holes (SMBHs) are common in galactic centers
and play an important role in galaxy formation and evolution \citep{Kormendy&Ho_araa}. 
However, the origin and early growth of central black holes (BHs) remain poorly understood.
In the current paradigm, 
SMBHs are thought to originate from seed BHs that formed at early cosmic time,
either through the evolution of Population III stars or 
via direct collapse of massive gas clouds \citep[e.g.,][]{Inayoshi2020}.
Intermediate-mass black holes (IMBHs), with masses \mbh $\approx 10^3-10^6$ \msun, 
are widely regarded as crucial links between BH seeds and SMBHs,
and may hold vital clues to understanding the seeding process.

Recent observations by the James Webb Space Telescope (JWST) have revealed 
BHs with masses exceeding $10^6$ \msun\ already in place at $z>10$ \citep{GN-z11,highz-agn}, 
only a few hundred million years after the Big Bang.
These discoveries indicate that rapid early BH growth is required 
and may favor scenarios involving massive initial seeds or highly efficient accretion.
However, direct detection of BH seeds at early cosmic epochs remains challenging,
and comprehensive surveys of high-redshift IMBHs are still limited. 
As a complementary approach, the census of low-redshift IMBHs may provide a valuable alternative approach to 
constrain BH formation theories and BH growth \citep[e.g.,][]{Inayoshi2020,greene_araa,Volonteri2021}.

This approach is also particularly important in the context of BH evolution at late cosmic times.
The growth of massive SMBHs after cosmic noon ($z\sim2-3$) has been extensively characterized through active galactic nucleus (AGN) 
luminosity function and BH mass function studies,
which reveal luminosity- and mass- dependent evolution commonly 
described as cosmic downsizing \citep[e.g.,][]{Ueda2003,Hasinger2005,merloni2008,shen&kelly2012}.
However, these observational constraints primarily probe BHs above $\sim 10^{8}$ \msun, 
leaving the evolutionary behavior of lower-mass systems largely unconstrained.

Attempts to probe the low-mass end of the active BH population 
have begun to provide clues to late-time growth. 
Observational studies suggest that active IMBHs in the low-redshift Universe appear to exhibit 
relatively strong accretion activity compared with their more massive counterparts,
with a comoving volume density exceeding that of massive SMBHs \citep{heckman04,gh07MF,Cho&WooMF2024},
implying that significant BH growth may continue well after cosmic noon.
Nevertheless, these studies mainly describe demographic properties within limited redshift intervals 
and therefore represent population snapshots rather than direct evolutionary measurements,
and the growth history of IMBHs at late cosmic times remains observationally unclear.

On the theoretical side, predictions for the evolution of low-mass BHs and IMBHs diverge markedly.
These differences largely arise from distinct assumptions regarding BH seeding efficiency, 
gas fueling channels, and the treatment of stellar and AGN feedback.
Some semi-analytic models predict substantial low-redshift growth in low-mass ($<10^7$ \msun) BHs,
driven in part by mass-dependent Eddington ratio distributions \citep[e.g.,][]{shankar2013,Fontanot2020}.
In contrast, population-synthesis models \citep{Fanidakis2012} predict active BH number densities that peak at high redshift ($z\approx3.5$) and decline steadily toward the present epoch.
Hydrodynamical simulations further illustrate the sensitivity of low-mass BH evolution 
to feedback and gas regulation.
For example, \citet{Hirschmann2014} reproduce both the mass- and luminosity-dependent downsizing 
pattern of AGN activity after $z<2$ within a cosmological framework,
although their simulation significantly underpredicts the abundance of BHs below 
$\sim5\times10^{7},\mathrm{M_\odot}$ at $z\approx0$ compared to the observed BH mass function,
largely owing to limitations in seed mass prescriptions and numerical resolution.
Within this framework, the downsizing behavior is attributed to the declining gas supply 
in the vicinity of BHs as a consequence of star formation and AGN feedback, 
and the resulting accretion histories give rise to a mass-dependent evolution of the 
Eddington-ratio distribution.
In contrast, the Illustris simulation \citep{sijacki2015} shows a decline in AGN comoving 
number density at $z<2$ without a delayed peak at low luminosities, 
implying different evolutionary behavior for low-mass systems;
in this framework, the evolution of the Eddington-ratio distribution emerges independent of BH mass,
These contrasting results highlight how different assumptions about gas fueling, feedback, and accretion prescriptions lead to markedly different expectations for the evolution of low-mass AGNs, underscoring the need for empirical constraints from well-defined IMBH samples.

Motivated by these open questions, 
considerable effort has been devoted to identifying accreting IMBHs in the nearby Universe.
The identification of IMBHs presents significant observational challenges, even at low redshifts,
primarily due to their intrinsically low luminosities.
Moreover, emission from the star-forming regions in the late-type host galaxies of IMBHs 
often obscures and severely contaminates the signatures of AGNs,
further complicating the detection.
Despite these challenges, targeted studies over the last two decades have successfully 
identified a growing number of low-mass AGNs. 
One particularly effective approach involves detecting AGNs through broad emission lines,
such as \ha\ and \hb, which trace high-velocity gas within the broad-line region surrounding the BH.
The widths of these lines, combined with the nuclear luminosity, enable BH mass estimation, 
providing a direct diagnostic of BH mass. 
The studies by \citet{NGC4395_Filippenko2003}, \citet{Barth2004}, and \citet{gh04,gh07} first 
demonstrated the efficacy of using the broad \ha\ line to identify IMBHs, 
establishing a robust sample of accreting BHs with masses below $2\times10^{6}$ \msun. 
Subsequent studies have expanded these efforts using larger spectroscopic surveys, 
uncovering hundreds of accreting IMBHs \citep[e.g.,][]{dong12,Reines2013,liu18,Pucha2025}.

In this study, we present a new systematic search for IMBH AGNs utilizing 
spectra from the Seventeenth Data Release \citep[DR17;][]{SDSSDR17} of the Sloan Digital Sky Survey (SDSS).
Our sample is primarily selected based on the presence of broad \ha\ emission in optical spectra,
employing an improved broad-line selection process designed to enhance the identification efficiency of 
AGN candidates, thereby improving the completeness of the sample.
By making full use of the wide wavelength coverage of 
Baryon Oscillation Spectroscopic Survey/extended Baryon Oscillation Spectroscopic Survey (BOSS/eBOSS)\footnote{BOSS and eBOSS are part of SDSS-III and SDSS-IV, respectively.} spectra, 
we have identified 930 sources and extend the sample's redshift range up to $z=0.57$---a marked improvement over previous studies typically limited to $z\lesssim0.35$ \citep{gh07,dong12,liu18}.
This expanded redshift range and increased sample size enable more robust statistical analyses
of the demographic properties of IMBH AGNs across a broader cosmic timescale.

This paper is structured as follows: 
Section~2 describes the spectroscopic dataset used in this work.
Section~3 introduces the spectral fitting methodology and the sample selection in detail.
In Section~4, the sample properties are discussed. 
Finally, we summarize our main results in Section~5.
Throughout this paper, we adopt a cosmology with $H_{0}=70$\,\kms\,Mpc$^{-1}$, $\Omega_\mathrm{m}=0.3$, and $\Omega_{\Lambda}=0.7$.

\section{Data}\label{section:data}

The IMBH sample is uniformly selected from the SDSS DR17 spectroscopic dataset,
which incorporates data from SDSS-I,II/SDSS Legacy \citep{SDSSDR7}, SDSS-III/BOSS \citep{BOSS},
and SDSS-IV/eBOSS \citep{eBOSS} surveys. 
All spectra were obtained with a resolving power of $\lambda/\Delta \lambda\approx2000$
and an instrumental dispersion of $\sim69$ \kms.
The diameters of the fibers used in SDSS are 3\arcsec\ for SDSS-I,II/Legacy
and 2\arcsec\ for SDSS-III/BOSS and SDSS-IV/eBOSS, respectively.
We include all DR17 spectra classified either as ``GALAXY'' or ``QSO'' by the SDSS pipeline.

To identify broad-line AGNs through their \ha\ or \hb\ emission, 
we require these lines to lie within the SDSS spectral coverage.
The wavelength range of the SDSS Legacy spectra is 3800--9200\,\AA, 
while BOSS/eBOSS extends from 3650 to 10400\,\AA. 
Accordingly, we select SDSS Legacy spectra with $z\leqslant0.35$,
for which the broad \ha\ line falls within the observed wavelength range.
Owing to the more limited wavelength coverage of the SDSS Legacy spectra,
\ha\ is not accessible at $z>0.35$.
To avoid incompleteness in this redshift range,
we additionally include SDSS Legacy objects at $0.35<z\leqslant 0.57$ and 
identify broad-line AGNs based on the presence of broad \hb\ emission.

After applying these redshift criteria, 
our dataset includes 1,009,616 SDSS Legacy spectra (976,929 classified as ``GALAXY'' and 32,687 classified as ``QSO'') 
and 1,365,332 BOSS/eBOSS spectra (1,284,764 ``GALAXY'' and 80,568 ``QSO''), 
with duplicate observations retained.
All spectra are corrected for Galactic extinction using the dust map of \citet{Schlegel1998}
and the reddening curve of \citet{Fitzpatrick1999}.
Each spectrum is then shifted to the rest frame using the redshifts provided by the SDSS pipeline.

For sources with multiple SDSS observations,
we stack the individual spectra to obtain a single, high signal-to-noise ratio (S/N) spectrum,
using an inverse-variance-weighted mean after bringing the spectra to a common wavelength grid.
All spectral fitting, as detailed in the next section, is performed on the stacked spectra,
and the number of available spectra per source is listed in Table~\ref{table:objinfo}.

\begin{figure*}[htbp]
   \centering
   \includegraphics[width=0.58\textwidth]{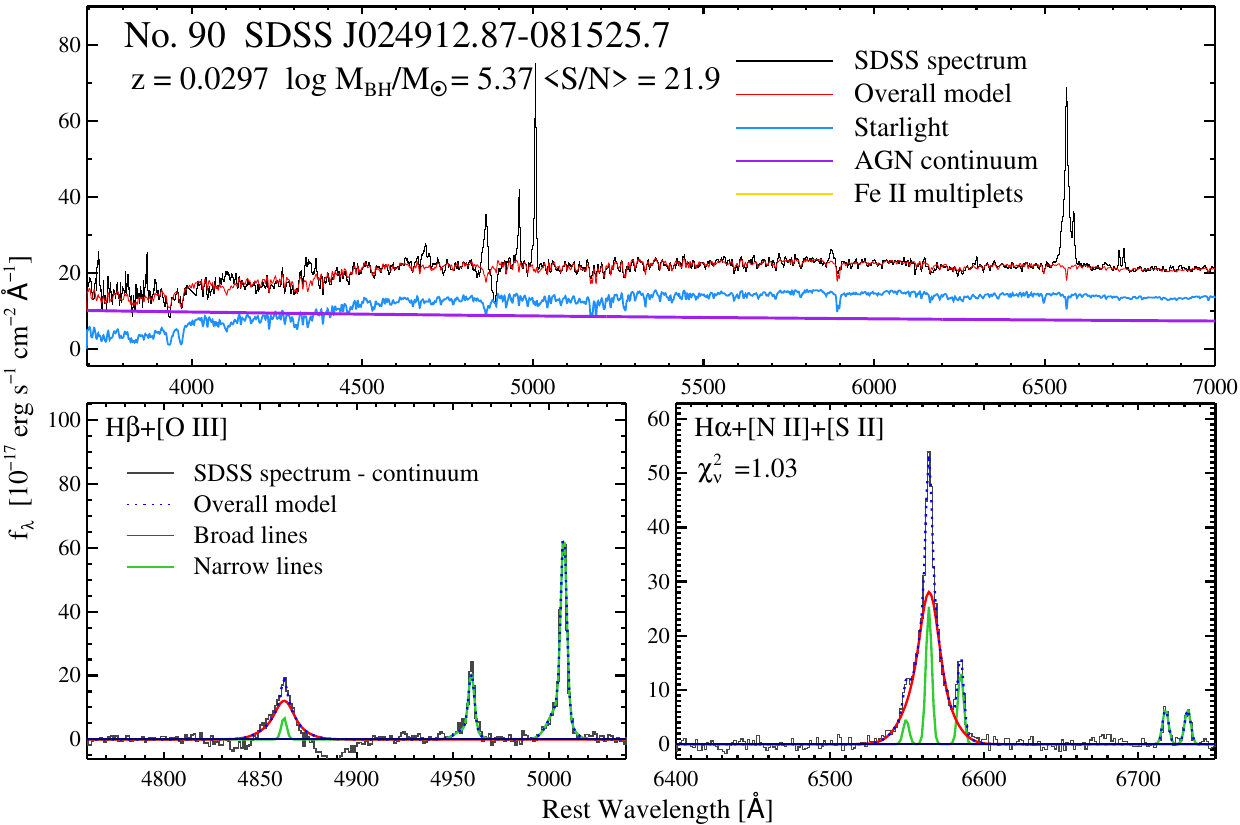}
   \includegraphics[width=0.58\textwidth]{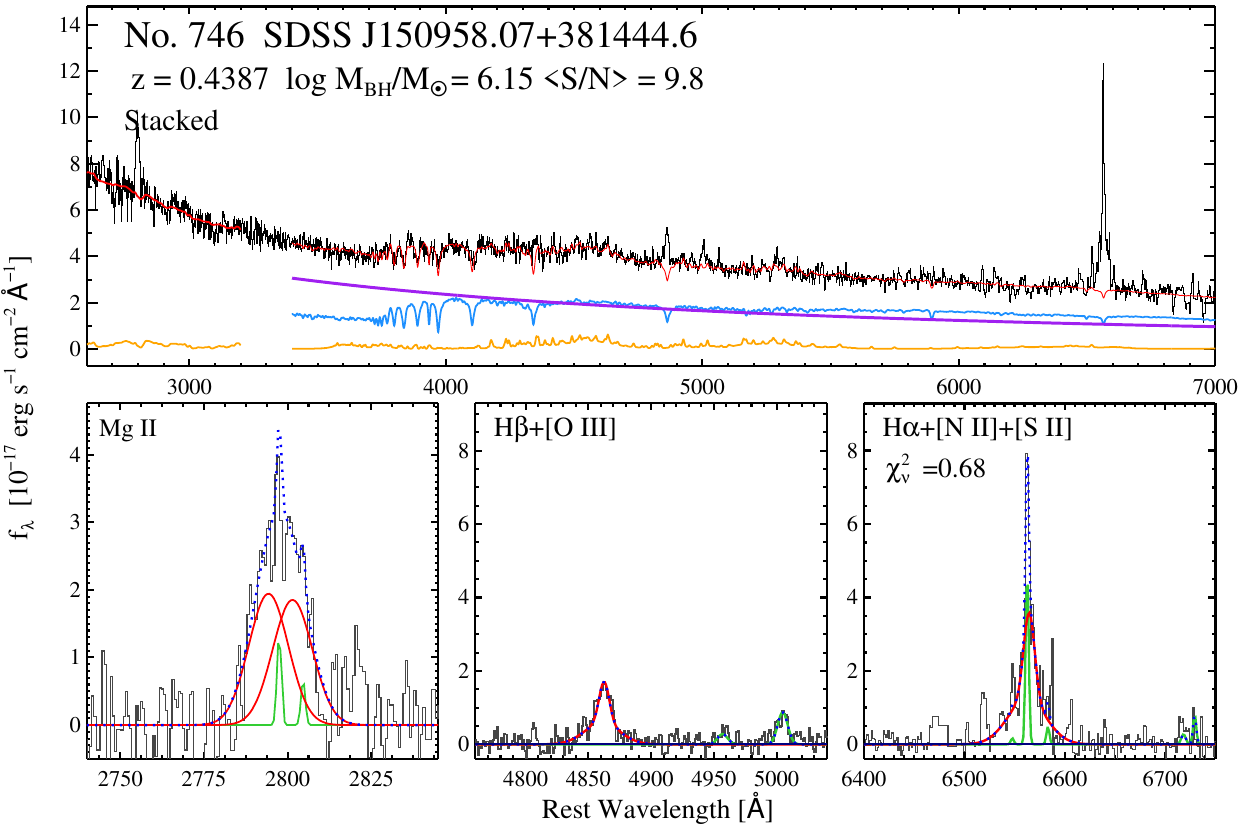}
   \includegraphics[width=0.58\textwidth]{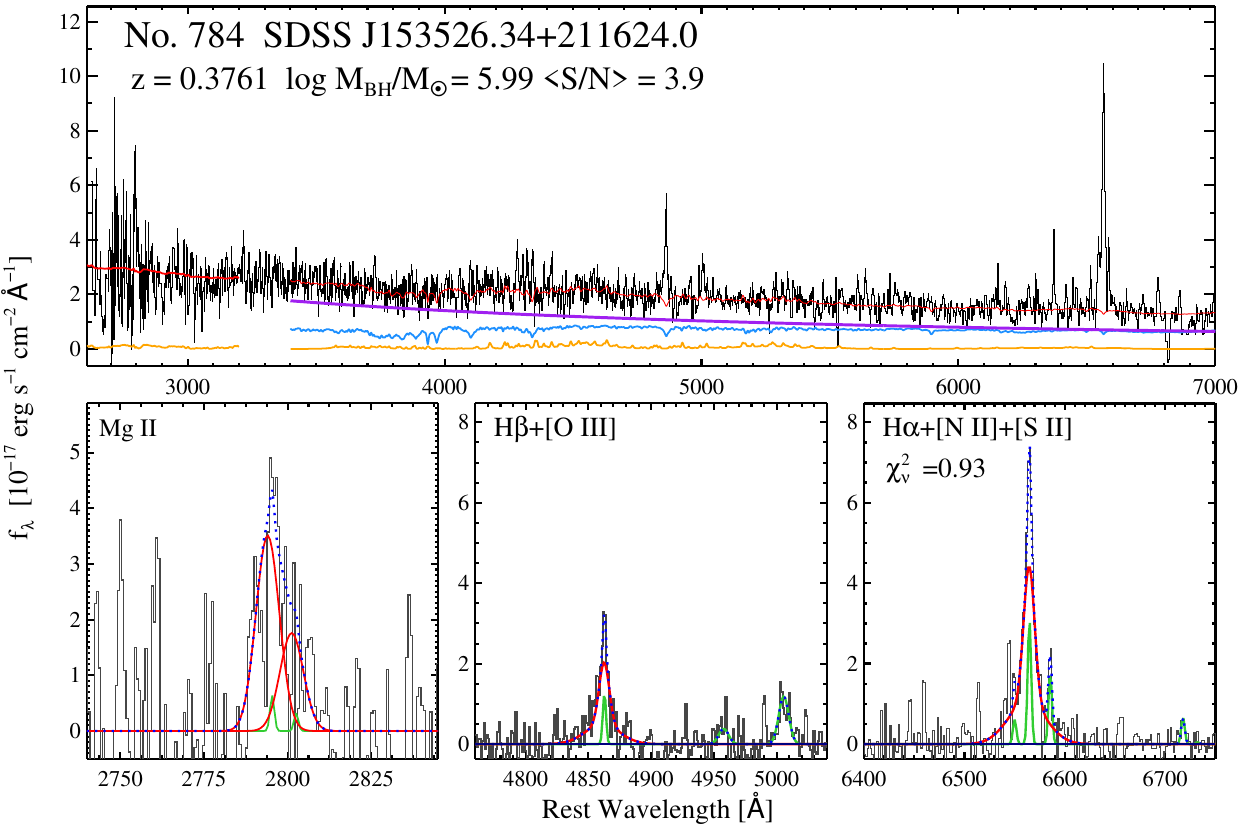}
   \caption{Illustration of the continuum and emission-line fitting for three example IMBH AGNs. 
   For each object, the upper panel shows the continuum decomposition based on the SDSS spectrum (black), 
   with the total model overlaid in red. 
   The AGN power-law continuum, host galaxy starlight, and \feii\ multiplets (optical and UV) are shown 
   in purple, blue, and orange, respectively. 
   In the \mgii\ region, 
   due to large uncertainty in UV continuum decomposition (see \S~\ref{subsec:mg2fit}), 
   only the total continuum model and the UV \feii\ component are displayed.
   For clarity, the SDSS spectra are smoothed using a 3 pixel boxcar filter.
   The lower panels display the emission-line fitting for the \mgii, \hb+\oiii\ and \ha+\nii+\sii\ regions.
   The continuum-subtracted spectrum is shown in black,
   the total emission-line fits in blue dotted lines,
   and the broad and narrow components in red and green, respectively.  
   No spectral smoothing was applied in the emission-line panels.
   The figures for the continuum and emission-line decomposition for all 930 sources are available in the journal's online supplementary files.}\label{fig:fig1}
\end{figure*}

\section{Data Analysis and Sample Selection}

\subsection{Spectral Fitting and Selection Criteria}

Our sample is primarily selected following the data analysis procedures outlined in \citet{dong12} and \citet{liu19}, 
with updates to improve the detection of weak broad lines in IMBH AGNs.
Spectral fitting is performed in IDL using the MPFIT package \citep{mpfit},
which applies $\chisq$ minimization via the Levenberg-Marquardt algorithm.
Details of the continuum and emission-line model configuration and the overall fitting procedures are described below.

\subsubsection{Continuum Model Configurations}

We model the continuum over the wavelength range 3700-7500\,\AA.
Due to the 3\arcsec\ or 2\arcsec\ diameter apertures used for SDSS spectroscopy, 
many spectra include significant contributions from host galaxy starlight. 
Accurate subtraction of the stellar continuum is therefore 
essential for reliable measurements of emission lines.

As the first step in the fitting process, 
we model the continuum using the method described in \citet{zhou2006} and \citet{dong12}.
The continuum is modeled as a linear combination of three primary components:
host galaxy starlight, AGN nuclear continuum, and optical \feii\ multiplets.
For starlight, 
we use six synthetic galaxy templates from \citet{lu2006},
constructed via independent component analysis from the simple stellar population models of \citet{BC03}.
These templates effectively capture a wide range of stellar features 
while mitigating the overfitting problem.
During fitting, the templates are convolved with a Gaussian kernel 
to match the widths of the stellar absorption features,
with the Gaussian width determined through a grid-based \chisq\ minimization up to 500 \kms, 
which safely exceeds the stellar velocity dispersions expected for galaxies hosting broad-line AGNs \citep{Kormendy&Ho_araa}.
We also allow small velocity shifts around the SDSS pipeline value and adopt the solution that minimizes the $\chisq$.  
The AGN continuum is modeled with a single power law.

The optical \feii\ multiplets are modeled using two separate sets of analytical templates constructed by \citet{dong08,dong11}, 
based on the measurements of the narrow-line Seyfert galaxy I\,Zw\,1 \citep{veron04},
which is commonly adopted as a reference source due to its strong and well-resolved \feii\ multiplets.
One set is used to model broad \feii\ emission and the other to model narrow \feii.
In these two sets of templates, the relative intensities of individual \feii\ lines are fixed to those measured in I\,Zw\,1,
while the kinematic properties of the broad and narrow components are modeled independently
in a manner similar to, but not identical with, the implementation described in \citet{dong08}. 
Specifically, the broad and narrow \feii\ components are initialized following the widths 
of the broad and narrow \hb\ lines, respectively, 
but are allowed to vary within reasonable ranges during the fitting process.
This analytical formulation provides sufficient flexibility to accommodate \feii\ emission 
significantly narrower than that of I\,Zw\,1, as often observed in IMBH AGNs.


In 25 objects in the sample with high Eddington ratios,
the stellar contributions are minimal, 
and stellar absorption features (e.g., \caii\,K, \caii\,H+H$\epsilon$) are weak or nearly absent.
In such cases, including stellar templates often leads to unphysically large broadening (e.g., $>500$ \kms),
as the fit attempts to reproduce marginal stellar absorption features.
To avoid this, following \citet{dong08,dong12},
we model the continuum of these spectra using only the AGN power law and \feii\ templates.

\begin{deluxetable*}{ccccccc}
\centering
\tablewidth{0pt}
\tablecaption{The SDSS DR17 IMBH-AGN Sample \label{table:objinfo}}
\tablehead{
\colhead{ID}  & \colhead{Designation} & \colhead{$z$} & \colhead{MJD}& \colhead{Plate} & \colhead{FiberID} & \colhead{$N_\mathrm{spec}$} \\
\colhead{(1)} & \colhead{(2)} & \colhead{(3)} & \colhead{(4)} & \colhead{(5)} & (6) & (7) 
}
\startdata
  1   &  J00:00:10.92$+$30:06:08.3 & 0.2606 & 58073 & 7749 & 0375 &   1 \\    
  2   &  J00:00:48.78$-$07:09:11.7 & 0.0375 & 56591 & 7148 & 0518 &   1 \\   
  3   &  J00:01:11.15$-$10:01:55.5 & 0.0489 & 52143 & 0650 & 0174 &   1 \\   
  4   &  J00:01:59.77$+$06:59:26.4 & 0.3001 & 57367 & 8740 & 0785 &   2 \\   
  5   &  J00:03:08.47$+$15:48:42.4 & 0.1176 & 52235 & 0750 & 0609 &   1 \\   
  6   &  J00:04:52.00$+$00:08:40.6 & 0.1868 & 52203 & 0685 & 0579 &   1 \\            
  7   &  J00:06:05.59$-$09:20:07.0 & 0.0699 & 52141 & 0651 & 0434 &   1 \\           
  8   &  J00:15:27.96$-$01:43:03.0 & 0.2124 & 56975 & 7863 & 0067 &   1 \\            
  9   &  J00:16:41.40$+$03:24:19.2 & 0.5327 & 57391 & 8745 & 0300 &   1 \\            
  10  &  J00:17:04.87$-$00:29:36.1 & 0.0672 & 51816 & 0390 & 0263 &   1 \\          
\enddata
\vspace{0.5em} 
\textbf{Notes.}\\
Column (1): Identification number assigned in this paper. \\
Column (2): Official SDSS designation in J2000.\\
Column (3): Redshift measured by the SDSS pipeline.\\
Column (4): Spectroscopic plate number.\\
Column (5): MJD of spectroscopic observation.\\
Column (6): Spectroscopic fiber number.\\
Column (7): Number of spectroscopic observations in SDSS surveys. 
We stack sources with multiple spectra (N$_\mathrm{spec}>1$), 
and the measurements reported in Table~\ref{table:lineinfo} for these sources are based on the stacked spectra.\\
(This table is available in its entirety in a machine-readable form in the online journal. 
A portion is shown here for guidance regarding its form and content.)

\end{deluxetable*}

\begin{deluxetable*}{cRRRRRRRRRRRRRRRRRR} 
 \centerwidetable
 \movetableright -0.3in
 \setlength{\tabcolsep}{2.pt}
\rotate{}
\tablewidth{0pt}
\tabletypesize{\tiny}
\tablecaption{Emission-line Measurements \label{table:lineinfo}}
\tablehead{
\colhead{\fontsize{6pt}{7.2pt}\selectfont ID} & \multicolumn{13}{c}{{\sixpt log Flux} } & \multicolumn{5}{c}{\sixpt FWHM}  \\
\colhead{}  & \multicolumn{13}{c}{\sixpt (erg~s$^{-1}$~cm$^{-2}$)} & \multicolumn{5}{c}{\sixpt ({\kms})} \\ 
\cline{2-14} \\[-9pt] \cline{15-19} \\[-6pt] \cline{15-19} 
\colhead{ } & \colhead{\sixpt [O\,{\tiny II}] $\lambda3727$} & \colhead{\sixpt Fe\,{\tiny II}\,$\lambda4570$} & 
\colhead{\sixpt H$\beta^\mathrm{n}$} &  \colhead{\sixpt [O\,{\tiny III}] $\lambda5007$} & 
\colhead{\sixpt [O\,{\tiny I}] $\lambda6300$} & 
\colhead{\sixpt H$\alpha^\mathrm{n}$} & \colhead{\sixpt [N\,{\tiny II}] $\lambda6583$} & 
\colhead{\sixpt [S\,{\tiny II}] $\lambda6716$} & \colhead{\sixpt[S\,{\tiny II}] $\lambda6731$} & 
\colhead{\sixpt Mg\,{\tiny II}$^\mathrm{b}$} & \colhead{\sixpt H$\beta^\mathrm{b}$}  & 
\colhead{\sixpt H$\alpha^\mathrm{b}$} & \colhead{\sixpt $\sigma_\mathrm{t}\,(\ha^\mathrm{b})$} & 
\colhead{\sixpt [O\,{\tiny III}]} &  \colhead{\sixpt [S\,{\tiny II}]} &
\colhead{\sixpt Mg\,{\tiny II}$^\mathrm{b}$} & \colhead{\sixpt \ha$^\mathrm{b}$} & 
\colhead{\sixpt $\sigma_\mathrm{t}\,(\ha^\mathrm{b})$} \\
\colhead{\sixpt (1)} & \colhead{\sixpt (2)} & \colhead{\sixpt (3)} & 
\colhead{\sixpt (4)} & \colhead{\sixpt (5)} & \colhead{\sixpt (6)} &
\colhead{\sixpt (7)} & \colhead{\sixpt (8)} & \colhead{\sixpt (9)} & \colhead{\sixpt (10)} & 
\colhead{\sixpt (11)} & \colhead{\sixpt (12)} & \colhead{\sixpt (13)} & \colhead{\sixpt (14)} & 
\colhead{\sixpt (15)} & \colhead{\sixpt (16)} & \colhead{\sixpt (17)} &  \colhead{\sixpt (18)} & 
\colhead{\sixpt (19)}
}
\startdata
  1 & -15.20 $\pm$ 0.02 & -15.26 $\pm$ 0.01 & -15.47 $\pm$ 0.02 & -15.16 $\pm$ 0.02 & -16.11 $\pm$ 0.09 & -14.79 $\pm$ 0.01 & -15.31 $\pm$ 0.02 & -15.49 $\pm$ 0.04 & -15.71 $\pm$ 0.04 &  & <-15.13 & -14.69 $\pm$ 0.07 &  0.10 & 434$\pm$25 & 264$\pm$21 &  & 1066$\pm$31 & 225\\
  2 & -14.60 $\pm$ 0.03 & <-15.56 & -14.95 $\pm$ 0.01 & -13.81 $\pm$ 0.01 & -15.21 $\pm$ 0.02 & -14.29 $\pm$ 0.01 & -14.86 $\pm$ 0.01 & -14.94 $\pm$ 0.02 & -14.97 $\pm$ 0.02 &  & -14.83 $\pm$ 0.03 & -14.25 $\pm$ 0.05 &  0.12 & <156 & 196$\pm$8 &  & 1184$\pm$62 & 445\\
  3 & -14.82 $\pm$ 0.04 &  & -15.25 $\pm$ 0.03 & -14.65 $\pm$ 0.01 & -15.79 $\pm$ 0.06 & -14.53 $\pm$ 0.00 & -14.95 $\pm$ 0.01 & -15.18 $\pm$ 0.02 & -15.33 $\pm$ 0.02 &  & <-15.29 & -14.61 $\pm$ 0.08 &  0.21 & 213$\pm$13 & 196$\pm$13 &  & 1862$\pm$43 & 315\\
  4 & -16.18 $\pm$ 0.12 & <-15.84 & <-16.51 & -15.95 $\pm$ 0.02 & <-16.49 & -16.15 $\pm$ 0.05 & -16.27 $\pm$ 0.04 & -16.70 $\pm$ 0.01 & -16.70 $\pm$ 0.01 &  & -15.60 $\pm$ 0.02 & -15.09 $\pm$ 0.11 &  0.22 & 214$\pm$17 & 171$\pm$17 &  & 927$\pm$21 & 58\\
  5 & -15.20 $\pm$ 0.04 & -15.24 $\pm$ 0.02 & -15.58 $\pm$ 0.03 & -15.35 $\pm$ 0.03 & -16.32 $\pm$ 0.13 & -14.99 $\pm$ 0.02 & -15.39 $\pm$ 0.03 & -15.61 $\pm$ 0.03 & -15.68 $\pm$ 0.08 &  & -15.08 $\pm$ 0.02 & -14.57 $\pm$ 0.06 &  0.14 & 169$\pm$15 & <145 &  & 896$\pm$35 & 228\\
  6 & -15.97 $\pm$ 0.10 & -15.74 $\pm$ 0.02 & -15.85 $\pm$ 0.04 & -15.91 $\pm$ 0.07 & <-16.02 & -15.34 $\pm$ 0.03 & -15.77 $\pm$ 0.04 & -16.06 $\pm$ 0.10 & -15.80 $\pm$ 0.08 &  & -15.64 $\pm$ 0.05 & -15.09 $\pm$ 0.59 &  0.16 & 297$\pm$62 & 169$\pm$62 &  & 1338$\pm$108 & 333\\
  7 & -15.07 $\pm$ 0.05 & <-15.79 & -15.74 $\pm$ 0.04 & -14.78 $\pm$ 0.01 & -15.83 $\pm$ 0.07 & -14.96 $\pm$ 0.01 & -15.03 $\pm$ 0.01 & -15.45 $\pm$ 0.02 & -15.48 $\pm$ 0.02 &  & -15.03 $\pm$ 0.03 & -14.63 $\pm$ 0.12 &  0.12 & 199$\pm$10 & 175$\pm$10 &  & 1621$\pm$66 & 515\\
  8 & -15.77 $\pm$ 0.05 & -15.48 $\pm$ 0.01 & -15.91 $\pm$ 0.14 & -15.61 $\pm$ 0.03 & <-15.68 & -15.19 $\pm$ 0.05 & <-15.93 & <-16.12 & <-16.20 &  & -15.54 $\pm$ 0.08 & -14.88 $\pm$ 0.00 &  0.23 & 265$\pm$39 & 257$\pm$39 &  & 785$\pm$33 & 134\\
  9 & -16.28 $\pm$ 0.11 & -16.13 $\pm$ 0.02 & -16.46 $\pm$ 0.11 & -16.10 $\pm$ 0.06 &  & -15.97 $\pm$ 0.12 & -15.98 $\pm$ 0.09 & -16.63 $\pm$ 0.00 & <-15.17 & -15.37 $\pm$ 0.08 & <-15.69 & -15.46 $\pm$ 1.40 &  0.30 & 257$\pm$43 & 219$\pm$43 & 2464 & 976$\pm$111 & 116\\
  10 & -14.71 $\pm$ 0.06 &  & -14.67 $\pm$ 0.01 & -14.73 $\pm$ 0.02 & -15.57 $\pm$ 0.08 & -14.04 $\pm$ 0.01 & -14.38 $\pm$ 0.01 & -14.87 $\pm$ 0.02 & -14.98 $\pm$ 0.02 &  & <-14.91 & -14.34 $\pm$ 0.07 &  0.28 & 384$\pm$63 & 180$\pm$63 &  & 1708$\pm$74 & 289\\
\enddata
\vspace{0.5em}
\textbf{Notes:}
{\scriptsize
Column (1): Identification number assigned in this paper.
Columns (2)-(13): Emission-line fluxes. 
The listed uncertainties represent measurement errors obtained from spectral fitting.
Fluxes are observed values with no NLR or BLR extinction correction applied.
The superscripts ``n'' and ``b'' refer to the narrow and broad components of the line, respectively.
Mg\,{\scriptsize II}$^{b}$ flux in Col. (11) corresponds to the combined flux of the doublet 
Mg\,{\scriptsize II} $\lambda\lambda2796,2803$.
The measured emission-line fluxes are regarded to be reliable detections if they have significance greater than 3$\sigma$,
or else $3\sigma$ values are adopted as the upper limit.
Column (14):  Total uncertainty of the broad \ha\ flux, defined as $\sigma_\mathrm{t} = \sqrt{\sigma^2_\mathrm{m} + \sigma^2_\mathrm{sys}}$,
where $\sigma_\mathrm{m}$ is the formal measurement error from the best-fit model,
and $\sigma_\mathrm{sys}$ is a systematic uncertainty estimated from alternative narrow-line
decomposition schemes used to separate the broad and narrow \ha\ components. 
Columns (15)-(18): Line widths, corrected for instrumental broadening.
If the measured FWHM, after correcting for instrumental broadening, is smaller than the instrumental FWHM (2.35\,$\sigma_\mathrm{inst}$), 
its FWHM is reported as $<2.35\,\sigma_\mathrm{inst}$, indicating that the emission line is narrow and not well resolved.
Column (19): Total uncertainty of the broad \ha\ FWHM, 
computed in the same manner as in Col. (14).
(This table is available in its entirety in a machine-readable form in the online journal.
A portion is shown here for guidance regarding its form and content.)}
\end{deluxetable*}

\subsubsection{Emission-line Model Configurations \label{subsubsec:emissionlines}}

Our emission-line fitting aims to robustly identify and measure narrow and/or weak 
broad-line components.
We primarily focus on detecting broad \ha\ emission,
the most prominent broad line in the optical spectra of AGNs.
For a small subset of objects with insufficient spectral quality around \ha, 
we instead use \hb\ as a secondary diagnostic.

The fitting focuses on the \ha+\nii+\sii\ and \hb+\oiii\ spectral regions.
The \ha+\nii\ complex is generally challenging to model 
due to the blending of broad \ha\ with narrow \ha\ and the adjacent \nii\ doublet.
To accurately decompose broad \ha,
we use isolated, well-detected narrow lines as templates to constrain the profiles of \nii\ and narrow \ha.
When available, \sii\ is prioritized as the preferred narrow-line template.
Line widths of AGN forbidden lines are known to correlate with both critical density and ionization potential.
Both \sii\ and \nii\ are low-ionization species;
despite their significantly different critical densities,
\sii\ often has a line profile closely resembling that of \nii, 
as demonstrated in previous studies \citep[e.g.,][]{Ho1997b}.
Its proximity in wavelength to the \ha+\nii\ region also minimizes spectral resolution mismatch.
All of these make \sii\ both a physically and practically suitable template.
If \sii\ is unavailable or too noisy, we resort to \oiii$\,\lambda5007$ as a fallback.
Although \oiii\ is typically brighter than \sii,
its higher ionization potential and critical density often lead to broader and more asymmetric
profiles due to kinematic effects such as AGN-driven outflows.
These limitations should be kept in mind when using it as a template.

In general, our fitting configurations for emission lines are as follows:

{\bf Narrow-line modeling.}

(1) Generally, each narrow line is modeled with a single Gaussian by default.
When \sii\ or \oiii\ are well-detected (S/N$>10$) 
and display complex profiles that cannot be adequately represented by a single Gaussian, 
we introduced a second Gaussian.
The two-Gaussian scheme is only accepted if it statistically improves the fit by $P_{F-\mathrm{test}}<0.05$.
In the final sample, 56.5\% of the sources require a double-Gaussian model for \oiii,
while only 3.1\% require a double-Gaussian representation for \sii.

(2) For all sources,
the \oiii\,$\lambda\lambda4959,5007$ and \nii\,$\lambda\lambda6548,6583$ doublets
each originate from a common upper energy level within their respective pairs.
Accordingly, we fix the wavelength separation within each doublet to laboratory values and 
constrain the flux ratios to their theoretical values (2.98 for \oiii\ and 2.96 for \nii);
both lines in each doublet are assumed to share the same line profile.

(3) Although the \sii\,$\lambda\lambda6716,6731$ doublet 
originates from different fine-structure upper levels,
we generally assume a shared profile due to limited S/N in most SDSS spectra.
Independent profiles are fitted only when the data quality allows. 

(4) Narrow \ha\ and \nii\,$\lambda\lambda6548,6583$ 
are assumed to share the same velocity shift and line width 
-- a reasonable assumption for most AGNs, 
as supported by previous spectroscopic studies \citep[e.g.,][]{Ho1997b}. 

(5) The velocity shift and line widths of narrow \ha\ and \nii\ lines 
are constrained using \sii\ doublet as the primary template, 
with the core component of \oiii\ serving as an alternative reference when \sii\ is weak or unavailable.

(6) The line width and velocity shift of the narrow \hb\ are tied to those of the narrow \ha.
In cases where the narrow and broad \hb\ components can be independently separated,
we allow \hb\ to be fitted without such constraints.
In the final sample, 3.5\% sources have their narrow \hb\ component fitted independently.

{\bf Broad-line modeling.}

(1) Broad \ha\ and \hb\ components are modeled using one or two Gaussians.
Given that broad lines in IMBH AGNs are generally relatively narrow and/or weak,
one or two Gaussians are typically sufficient to reproduce their profiles.
We start with a single Gaussian and introduce a second one only when the fits is significantly improved ($P_{F-\mathrm{test}}<0.05$).
In the final sample, 82\% of the sources require a double-Gaussian model for the broad \ha\ component.

(2) The broad \hb\ is initially tied to broad \ha\ in velocity shift and width,
providing more robust constraints when the \hb\ line is too weak for independent modeling.
If this assumption limits the fit quality, 
we allow \hb\ to be fitted independently if warranted by the $F$-test ($P<0.05$).
About 2\% sources in the final sample require an independently fitted broad \hb\ component.

\subsubsection{Iterative Continuum and Emission-line Fitting Procedure \label{subsec:fittingprocedure}}

We adopt an iterative fitting strategy 
that alternates between continuum modeling and emission-line fitting, 
allowing each component to inform and refine the other.
Continuum fitting requires masking out emission lines,
while accurate emission-line modeling depends on reliable continuum subtraction,
making the two processes inherently interdependent.
To reconcile this, we begin with a continuum estimate, 
subtract it from the observed spectrum,
and fit the residual emission lines.
The resulting emission-line models provide improved definitions of the
wavelength ranges affected by Balmer and other emission lines, 
which are then masked in the subsequent continuum fitting, 
while the kinematic information derived from the Balmer lines is 
used to guide the initialization of the broad and narrow \feii\ components 
in the next continuum-fitting iteration. 
A cleaner continuum subtraction, in turn, 
leads to more reliable emission-line measurements. 
This cycle is repeated until both the continuum and emission-line fits 
converge to statistically and visually satisfactory solutions, 
which typically occurs within three to five iterations.

In practice, this iterative cycle is organized into three stages with increasing complexity, 
each designed to achieve specific modeling goals and potentially involving multiple iterations.

The full procedure unfolds in three stages with increasing complexity :

\textit{Stage 1.} 
We begin with two preliminary iterations using only stellar continuum models,
without including AGN-related components.
Emission lines are masked based on the composite QSO spectrum of \citet{vandenberk2001}.
At this stage, emission-line fitting is limited to narrow lines, excluding broad components.
Only spectra with narrow-line fluxes exceeding twice  
their measurement uncertainties are retained for further analysis.
This pre-selection state removes approximately 52\% of the parent sample, which do not exhibit significant emission lines. 

\textit{Stage 2.}
For spectra that pass the initial screening, 
we expand the continuum model to include power-law and \feii\ templates to account for potential AGN-related emission. 
Once the continuum fit reaches convergence (typically reduced $\chisq < 1.5$)
we assess the presence of broad components in \ha\ and \hb. 
Broad components are added if they significantly improve the fit,
as evaluated by an $F$-test with $P_{F-\mathrm{test}}<0.05$.
Narrow \ha\ and \nii\ line profiles are constrained using \sii\ or \oiii\ lines as described in \S~\ref{subsubsec:emissionlines}. 
After the emission-line fitting converges to a statistically acceptable solution ($\chisq < 1.5$),
we apply the following set of criteria to select robust broad-line AGN candidates:
$P_{F-\mathrm{test}} < 0.05$,
S/N$_{\ha^\mathrm{b}}$ $\geqslant 3$,
$h_\mathrm{B} \geqslant 2$\,rms, and 
FWHM$_{\ha^\mathrm{b}}$ $>$ FWHM$_{[\mathrm{O\,\text{\tiny III}]}\,\lambda5007}$.
Here, $P_{F-\mathrm{test}}$ is the statistical probability from the $F-$test that quantifies
the statistical significance of adding a broad component relative to a narrow-line-only model;
$h_\mathrm{B}$ is the peak height of the best-fit broad \ha\ profile;
and rms refers to the rms of the continuum-subtracted 
spectrum in the line-free region adjacent to \ha 
\footnote{
For the \ha\ region, the rms is measured from the continuum-subtracted 
spectrum within a nominal wavelength range of 6300-6750 \AA, 
excluding pixels within $\pm$2 times the FWHM of the broad \ha\ component from the line centroid. 
Narrow emission lines such as \oi$\lambda\lambda6300,6364$ and \sii$\lambda\lambda6716,6731$, 
are masked when present.
For broad-line AGNs with exceptionally broad \ha\ emission,
the wavelength window is expanded and verified by visual inspection 
to ensure that only line-free regions are used.
}.

The S/N$_{\ha^\mathrm{b}}$ is defined as the ratio of broad \ha\ flux to its total uncertainty
$\sigma_\mathrm{total}$:
\begin{equation*}
    \sigma^2_\mathrm{total} = \sigma^2_\mathrm{n} + \sigma^2_\mathrm{cont\_sub} + \sigma^2_\mathrm{NL\_sub},
\end{equation*}
where $\sigma^2_\mathrm{n}$ is the statistical uncertainty returned by the fitting code MPFIT,
$\sigma^2_\mathrm{cont\_sub}$ accounts for the uncertainty arising from continuum decomposition,
and $\sigma^2_\mathrm{NL\_sub}$ represents the uncertainty arising from narrow-line subtraction.

The term $\sigma_\mathrm{cont\_sub}$ is negligible and thus omitted from $\sigma_\mathrm{total}$ 
throughout the analysis in practice. 
Following \citet{dong12} and \citet{liu18},
we visually inspect the continuum subtraction to ensure that 
the associated uncertainty is well below the 1$\sigma$ spectral flux density error,
particularly for spectra with significant contributions from intermediate-age stellar populations.
By contrast, $\sigma_\mathrm{NL\_sub}$ can be nonnegligible, 
especially for relatively narrow or weak broad \ha\ lines, 
as it depends on the adopted narrow-line model.
However, since this term is quantified through comparisons among multiple narrow-line modeling schemes, 
it is evaluated only in Stage 3.
At Stage 2, to enable efficient screening of our large sample, 
we adopt a preliminary estimate of $\sigma_\mathrm{total} \approx \sigma_\mathrm{n}$, 
thereby applying intentionally lenient selection criteria to identify promising broad-line AGN candidates 
while minimizing unnecessary refinement of non-BLAGN spectra.

We do not impose an explicit lower limit on the reduced \chisq\ values of the spectral fits.
Instead, the adopted broad-line selection criteria -- in particular S/N$_{\ha^\mathrm{b}}$ $\geqslant 3$ 
and $h_\mathrm{B} \geqslant 2$\,rms -- require the broad component to be detected with sufficient statistical
significance relative to the local noise level.
Fits dominated by noise fluctuations or unstable decompositions therefore fail to satisfy these criteria 
and are excluded from the sample without applying an additional $\chisq$ threshold.
The resulting reduced \chisq\ values for the broad \ha\ fitting region span 0.62 -- 1.5 in the final sample,
consistent with expectations for SDSS spectral decompositions given the pipeline noise estimates.

\textit{Stage 3.}
The template-constrained fitting described above provides a physically motivated anchor 
for decomposing the blended broad and narrow components in the \ha+\nii\ complex. 
However, due to intrinsic differences in ionization potential and critical density among various AGN narrow lines,
subtle profile mismatches may exist between different narrow emission lines.
To incorporate these differences, 
we perform a refinement step for sources identified as broad-line AGNs
by testing multiple narrow-line modeling schemes.
These include both template-based approaches and fully unconstrained fits, 
allowing greater flexibility in modeling narrow-line components.
Specifically, we implement several schemes for modeling the narrow \ha\ profile,
including using the best-fit \sii\ profile (single or double Gaussian),
adopting either the core component or the full profile of \oiii, 
and performing a free fit without tying the line widths to \sii\ or \oiii.
The resulting fits are compared, 
and the final best-fit solution is selected based on the minimum \chisq, 
with additional visual inspection to ensure physical plausibility.
For reference, in the final sample, 
13.1\% of the sources are best fitted with the narrow-line profiles tied to the \sii\ core component (single Gaussian), 
1.0\% to the full \sii\ profile (double Gaussian), 
17.6\% to the \oiii\ core component (single Gaussian), 
1.9\% to the full \oiii\ profile (double Gaussian), 
and 66.4\% using fully unconstrained narrow-line fits.

After selecting the best-fit model, 
we estimate $\sigma_\mathrm{NL\_sub}$ from the dispersion among acceptable fitting solutions
obtained under different narrow-line modeling schemes,
following \citet{dong12} and \citet{liu18}:
\begin{equation*}
    \sigma_\mathrm{NL\_sub} = \sqrt{\frac{\sum^n_{i}(Flux(\ha^\mathrm{B})_{i}-Flux(\ha^\mathrm{B})_\mathrm{best})^2}{n}}
\end{equation*}
This allows us to compute the full uncertainty $\sigma_\mathrm{total}$.
For our sample, we find $\sigma_\mathrm{total} \approx 5\sigma_\mathrm{n}$ for the broad \ha\ flux.
Similarly, for the FWHM of broad \ha, the total uncertainty is $\sigma_\mathrm{total} \approx 4.2\sigma_\mathrm{n}$. 
These results are in good agreement with those reported by \citet{dong12}.
To ensure robustness, we then reapply the Stage 2 criteria using the refined model parameters 
and the updated (stricter) uncertainties, 
retaining only those sources that continue to meet all thresholds under this more rigorous evaluation.

\subsubsection{\mgii\ Fitting for Objects with $z>0.3$ \label{subsec:mg2fit}}

The extended wavelength coverage of BOSS/eBOSS spectra (3600\,--\,10400\,\AA)
enables the detection of \mgii\ emission in AGNs at $z\gtrsim0.3$.
We therefore perform a systematic analysis of the \mgii\ emission complex for sources in this redshift range.

The UV continuum underlying \mgii\ may include contributions from the AGN power-law emission,
Balmer continuum, and blended UV \feii\ multiplets. 
The Balmer continuum is largely featureless except for the 3646\AA\ jump,
making it strongly degenerate with the power-law component.
In IMBH AGNs, additional complications arise because the intrinsic AGN luminosity is low,
and the host-galaxy contribution around the Balmer jump can be significant.
Combined with the generally modest spectral quality of our \mgii\ accessible subsample 
(median S/N of 2.3-9.6 in \mgii\ region, typically $\sim 5$), 
these factors prevent a full physical decomposition of individual continuum components.
Under these conditions, 
our primary goal is to obtain a robust local continuum subtraction beneath \mgii\ rather 
than fully disentangle the physical contributions.

We adopt a fitting methodology adapted from \citet{wang09} for broad-line quasars, 
with modifications to account for the distinct properties of IMBHs, 
such as their lower luminosities and narrower emission-line widths.
Within this framework, the continuum is modeled over a restricted rest-frame window of 2700 -- 3300 \AA, 
which restricts the fit to a localized window, 
while providing sufficient wavelength coverage to characterize the local continuum around \mgii.
The continuum is represented using a combination of a power law and UV \feii\ multiplets. 
The power-law component effectively represents the overall featureless continuum, 
including any contribution from the Balmer continuum, 
while the UV \feii\ emission is modeled using the semi-empirical template from \citet{uvFeII_T06},
derived from I\,Zw\,1, broadened and velocity-shifted to match the observed profiles.

Each line of the \mgii\,$\lambda\lambda2796,2803$ doublet is modeled with broad and narrow components.
Both the broad and narrow components of the \mgii\ doublet are assumed to share the same profile within each pair,
with fixed doublet separation and a flux ratio between 1:1 and 2:1 \citep{Laor1997}.
The broad component is fitted with one or two Gaussians, depending on the $F-$test results.

We detect broad \mgii\ lines in 24 IMBH AGNs at $z>0.3$, 
as summarized in Table~\ref{table:lineinfo}. 
Example \mgii\ line fits are shown in Figure~\ref{fig:fig1}.

\subsection{The Final Sample of Broad-line IMBHs}

\begin{deluxetable}{CCCCC}
 \centerwidetable
 \movetableright=0in
\tablewidth{0pt}
\tabletypesize{\scriptsize}
\tablecaption{Physical Properties of the Sample\label{table:massinfo}}
\tablehead{
\colhead{ID} & \colhead{log\,$L_{\ha^\mathrm{b}}$} & \colhead{log\,\mbh$^{1}$ } & \colhead{log\,\redd$^{1}$} & \colhead{log\,\mbh$^{2}$} \\ 
\colhead{}   & \colhead{(erg~s$^{-1}$)} & \colhead{(\msun)} & \colhead{} & \colhead{(\msun)} \\
\colhead{(1)} & \colhead{(2)} & \colhead{(3)} & \colhead{(4)} & \colhead{(5)} 
}
\startdata
 1 & 41.63$\pm$ 0.04 & 6.29$\pm$ 0.19 & -0.44$\pm$ 0.19 & 6.52$\pm$ 0.19\\
 2 & 40.26$\pm$ 0.05 & 5.77$\pm$ 0.34 & -1.00$\pm$ 0.34 & 5.99$\pm$ 0.34\\
 3 & 40.15$\pm$ 0.09 & 6.12$\pm$ 0.16 & -1.44$\pm$ 0.17 & 6.34$\pm$ 0.16\\
 4 & 41.37$\pm$ 0.10 & 6.05$\pm$ 0.07 & -0.41$\pm$ 0.10 & 6.28$\pm$ 0.07\\
 5 & 40.99$\pm$ 0.06 & 5.85$\pm$ 0.23 & -0.50$\pm$ 0.23 & 6.07$\pm$ 0.23\\
 6 & 40.90$\pm$ 0.07 & 6.17$\pm$ 0.23 & -0.89$\pm$ 0.23 & 6.39$\pm$ 0.23\\
 7 & 40.45$\pm$ 0.05 & 6.13$\pm$ 0.29 & -1.22$\pm$ 0.29 & 6.35$\pm$ 0.29\\
 8 & 41.24$\pm$ 0.10 & 5.84$\pm$ 0.16 & -0.30$\pm$ 0.18 & 6.07$\pm$ 0.16\\
 9 & 41.59$\pm$ 0.13 & 6.19$\pm$ 0.12 & -0.38$\pm$ 0.16 & 6.43$\pm$ 0.12\\
10 & 40.69$\pm$ 0.12 & 6.29$\pm$ 0.16 & -1.18$\pm$ 0.19 & 6.52$\pm$ 0.16\\
\enddata
\vspace{0.5em}
{\textbf {\footnotesize Notes:}}
{
Column (1): Identification number assigned in this paper (an asterisk indicates sources with \mbh\ estimated from broad \hb; a total of seven sources.)
Column (2): Luminosity of broad \ha.
Column (3): Virial BH mass estimate using the broad \ha-based calibration from \citet{xiao11}.
Column (4): Eddington ratio.
Column (5): Virial BH mass estimate using the calibration from \citet{Ho_Kim2015}, applicable to host galaxies with either classical bulges or pseudobulges. 
The quoted uncertainties include measurement uncertainties propagated from the emission-line fitting, including systematic components associated with spectral decomposition. They do not include the intrinsic scatter of the adopted virial scaling relations (typically $\sim$ 0.4--0.5 dex). 
}
\end{deluxetable}

Applying the broad-line selection criteria to the SDSS spectroscopic dataset,
we construct a parent sample of $\sim$ 45,000 broad-line AGNs at $z<0.57$.
This full set of broad-line AGNs defines our parent sample 
and provides the context for the IMBH-AGN selection described below. The final IMBH-AGN sample comprises 930 sources.

Direct BH mass measurements via reverberation mapping are unfeasible for most broad-line AGNs due to observational constraints. 
Consequently, single-epoch spectral analysis has become the primary method for estimating BH masses,
especially for studies involving large AGN samples.
This approach involves measuring the width of broad emission lines and the continuum luminosity,
and applying the virial mass formalism based on the luminosity-broad-line region size relation calibrated through reverberation mapping.

For IMBHs, directly measuring nuclear continuum luminosities often proves challenging 
due to significant starlight contamination in their optical spectra.
We therefore adopt the broad \ha-based single 
BH mass estimation formalism from 
\cite[][hereafter X11; their Equation 6]{xiao11}.

This estimator is based on the \ha-based virial mass formalism developed by \citet{gh05},
in which BH masses are inferred from the broad \ha\ luminosity and 
line width through empirical scaling relations.
Subsequent work refined the calibration through 
updates to the reverberation-mapping radius-luminosity ($R-L$) relation.
In particular, \citet{gh07} implemented a revised $R-L$ relation,
and X11 further updated the calibration using the $R-L$ relation of \citet{bentz09}, yielding BH masses systematically lower by $\approx$0.08 dex relative to \citet{gh07}.
We adopt the X11 calibration to maintain consistency with previous optically selected IMBH samples \citep{gh07,dong12,liu18}.

More recent calibrations by \citet{Ho&Kim2014,Ho_Kim2015} offer further refinements 
by incorporating an improved $R-L$ relation and explicitly accounting for the differences between pseudo and classical bulges.
For completeness, we provide BH mass estimates based on both the X11 and 
\citet[][hereafter HK15]{Ho_Kim2015} formalisms in Table~\ref{table:massinfo}.
To ensure consistency with the \ha-based mass scale, we convert the HK15 \hb-based calibration to \ha\ using the \lfive--$L_{\ha}$ and $\mathrm{FWHM}_{\ha}-\mathrm{FWHM}_{\hb}$ relations established by \citet{gh05}.
The HK15 calibration yields systematically higher masses,
with a mean offset of $\Delta \mbh = \mathrm{log} \mbh (\mathrm{HK15}) - \mathrm{log} \mbh (\mathrm{X11}) = 0.23$\,dex.
Since our sample selection is based on the X11 mass estimates, 
all subsequent analysis is conducted using this calibration.

The rms scatter in the \lfive--$L_{\ha}$ relation is $\sim$0.2 dex \citep{gh05}, 
leading to typical formal measurement uncertainties of $\sim$0.14 dex in single-epoch BH masses
when propagating line luminosity and width errors (X11).
These formal uncertainties do not include the dominant systematic uncertainty associated 
with the virial normalization factor $f$, which contributes to the intrinsic scatter 
($\sim$0.4--0.5 dex) of single-epoch virial BH mass estimators. 

Our approach follows the standard framework adopted in previous optically selected IMBH studies \citep{gh07,xiao11,dong12,liu18}, in which single-epoch virial BH mass estimators are calibrated using reverberation-mapped AGNs primarily in the SMBH regime.
Its application to IMBHs therefore involves an extrapolation, which is unavoidable given the current lack of systematic reverberation-mapping constraints at low BH masses. Ongoing efforts to better map the IMBH parameter space will provide an independent assessment of these calibrations \citep[e.g.,][]{Sun2025}.

To assess the reliability of estimating continuum luminosities from broad \ha, 
we compare \lfive\ inferred from the \lfive--$L_{\ha}$ relation with directly measured 
continuum luminosities for sources whose spectra are dominated by AGN emission. 
Specifically, in the final sample, 195 objects have a power-law component 
contributing more than 70\% of the total continuum flux at 4200\,\AA, 
a wavelength region relatively free of strong emission lines in typical AGN spectra \citep[e.g.,][]{vandenberk2001},
allowing a reliable assessment of the continuum decomposition.
Empirical tests performed during the development of the fitting procedure show 
that above this threshold ($\geqslant$ 70\%) the recovered power-law component 
remains stable against variations in the fitting process. 
We find that the two luminosity estimates are consistent within a scatter of 0.23 dex, 
with a median offset of 0.14 dex. 
The resulting BH masses derived from the two approaches differ by 0.12 dex on average.
The modest systematic offset likely arises from 
residual uncertainties in continuum and emission-line decomposition, 
which can slightly bias the recovered AGN continuum level. 
Notably, the observed scatter is comparable to the $\sim$0.2 dex dispersion reported for the 
\lfive--$L_{\ha}$ relation in \citet{gh05},
indicating that our measurements are consistent with the
intrinsic level of uncertainty expected for \ha-based estimators.
These results support the applicability of the \ha-based method for the majority
of the sample, where direct continuum measurements are unreliable.




For the seven sources without reliable broad \ha\ measurements,
we infer the FWHM and luminosity of \ha\ from the corresponding broad \hb\ properties 
using the scaling relations from \citet{gh05}.
The resulting parameters are then used to estimate BH masses.

We compute the Eddington ratio (\redd), 
defined as the ratio of bolometric luminosity (\lbol) to the Eddington luminosity (\ledd),
where $\ledd=1.26\times10^{38} \mbh/\msun$ \ergs.
Bolometric luminosity (\lbol) is estimated from the monochromatic continuum luminosity at 5100\,\AA\ 
using the conversion provided by \citet{runnoe12}:
$\log \lbol = (4.891\pm1.657) + (0.912\pm0.037) \log \lambda L_{5100} + \log f$,
where $f$ is a dimensionless correction factor accounting for the average AGN viewing angle and the anisotropic emission of the accretion disk, and is approximated as 0.75 following \citet{runnoe12}.
In this study, we derive $\log L_{5100}$ according to the empirical correlations of 
$\lambda L_{5100}-L_{\ha}$ or $\lambda L_{5100}-L_{\hb}$ from
\citet{gh05}, given the challenges in directly measuring $\lambda L_{5100}$ for
starlight-contaminated spectra.

Applying the BH mass threshold of $\mbh \leqslant2\times10^{6}$ \msun, 
as introduced by \citet{gh07}, 
we compile a final sample of 930 IMBH AGNs.
Basic source information is provided in Table~\ref{table:objinfo},
while Table~\ref{table:lineinfo} and \ref{table:massinfo} 
summarize the emission-line measured parameters and derived physical properties (including luminosities of broad \ha, BH masses, and Eddington ratios), respectively.
In Table~\ref{table:lineinfo}, quoted uncertainties of emission-line parameters represent measurement errors
from spectral fitting, while for the broad \ha\ component,
we additionally report total uncertainties that include systematic uncertainties associated with narrow-line decomposition. 
The uncertainties listed in Table~\ref{table:massinfo} are propagated from these measurements and therefore
include both measurement and decomposition-related systematic components,
but do not include the intrinsic scatter of the adopted empirical scaling relations.
Among the sample, 96 objects have multiple SDSS spectra.
In the catalog (Table~\ref{table:objinfo}),
we record the number of SDSS spectra available for each source ($N_\mathrm{spec}$).
Spectral measurements for sources with $N_\mathrm{spec} > 1$ are based on stacked spectra, 
as described in Section~\ref{section:data}.

Figure~\ref{fig:fig1} illustrates representative examples of the spectral decomposition.
Panel (a) shows an object (SDSS J0249$-$0815) with a relatively low BH mass and moderate spectral quality.
Panels (b) and (c) present two IMBH candidates (SDSS J1509$+$3814, J1535$+$2116) at $z>0.35$ with detectable \mgii\ emission.
In particular, the spectrum in panel (c) has comparatively low overall data quality, 
yet a broad \ha\ component remains clearly identifiable, 
demonstrating the capability of our fitting procedure to recover broad lines even in low-quality spectra.

\subsection{Comparison with Prior IMBH Samples}

\begin{figure}[htbp]%
	\centering
        \includegraphics[width=0.495\textwidth]{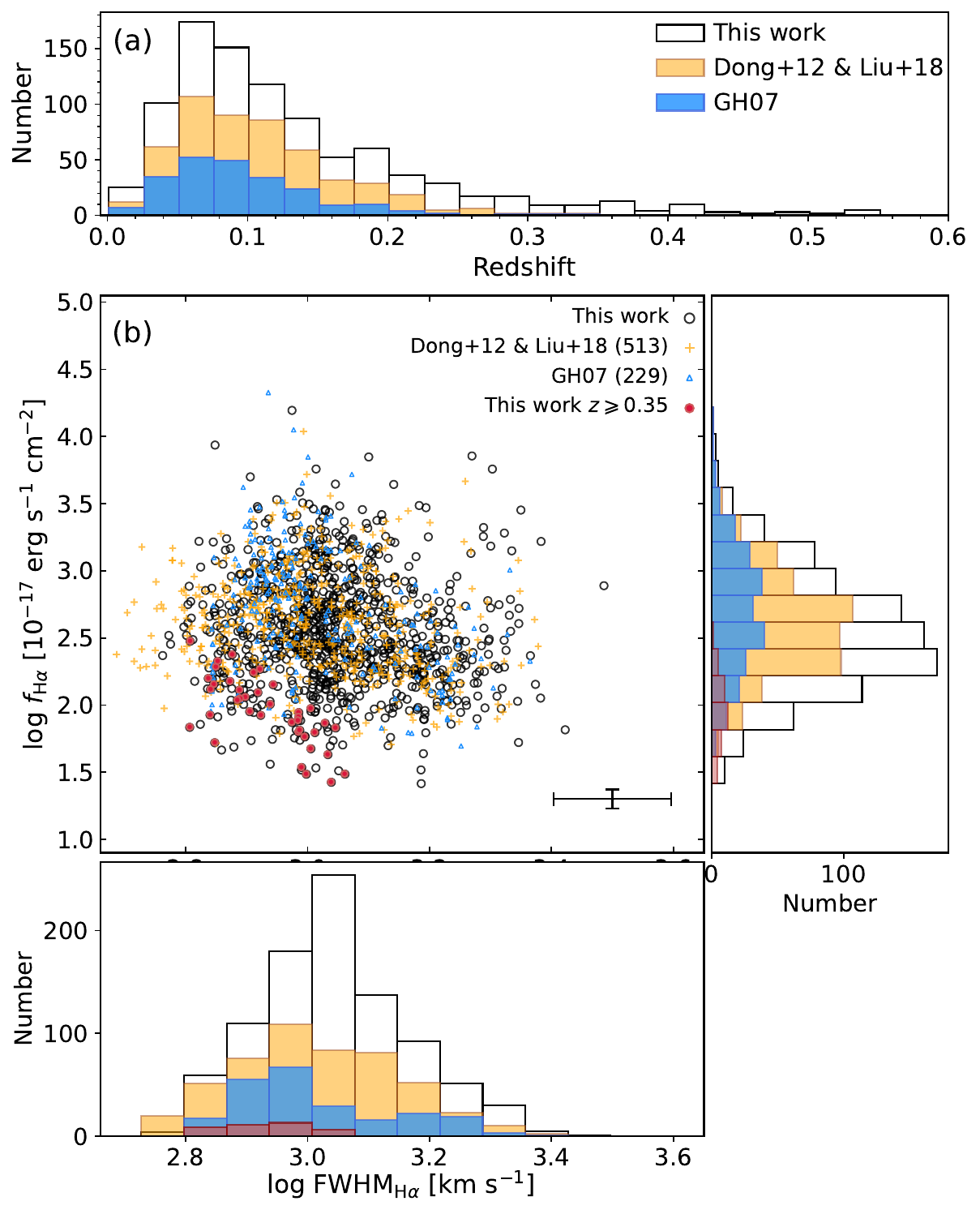}
	\caption{(a) Redshift distribution for our IMBH sample, 
                 compared with that from \citet{gh07}, \citet{dong12}, and \citet{liu18}.
             (b) Distribution of broad \ha\ of our IMBH sample at $z\leqslant 0.57$,
                 with $z\geqslant 0.35$ objects highlighted in red.
                 For comparison, the IMBH samples from \citet{gh07}, 
                 \citet{dong12}, and \citet{liu18} are displayed.
                 A typical 1-$\sigma$ error bar is shown (lower-right), 
                 representing the total uncertainty ($\sigma_\mathrm{total}$; see \citealt{dong12}, \S~2.3),
                 which combines contributions from statistical noise ($\sigma_\mathrm{n}$) and subtraction of nearby narrow lines ($\sigma_{\mathrm{NL\_sub}}$).
                 For our sample, $\sigma_\mathrm{total} \approx 5\,\sigma_\mathrm{n}$ for flux 
                 and $\approx 4.2\,\sigma_\mathrm{n}$ for FWHM, consistent with \citet{dong12}.
            }
	\label{fig:fig2}
\end{figure}

Earlier studies using SDSS spectroscopy \citep{gh07,dong12,liu18,chilingarian18}
have built considerable samples of low-$z$ IMBHs in broad-line AGNs.
More recently, \citet{Pucha2025} utilized DESI Early Data Release (EDR) spectra to construct 
another large IMBH candidate sample. 
Those efforts laid important groundwork for IMBH searches using single-epoch spectroscopy.
Here, we compare our newly constructed catalog with those previous works to assess the 
overlap and evaluate improvements in detection sensitivity.

Our sample recovers 459 of 513 sources identified by \citet{dong12} and \citet{liu18},
and 199 of the 229 sources from the sample of \citet{gh07}.
Among the unmatched objects, many remain in our broad-line AGN sample but exceed the IMBH mass threshold based on our line decomposition;
specifically, 48 sources from the former and 20 from the latter fall into this category. 
The remaining six and 10 sources, respectively, 
are excluded altogether based on our emission-line decomposition and/or selection criteria. 
A key reason for exclusion is that these sources exhibit relatively broad, narrow lines, including the more isolated lines such as \sii\ and \oi. In such objects, their narrow-line profiles in the \ha+\nii\ region can be well modeled using the global \sii\ or \oiii\ templates without the need for an additional broad component.


We also compare with the IMBH sample compiled by \citet{chilingarian18},
who identified 305 candidates based on SDSS DR7 spectra with 
$3\times10^4 < \mbh < 2\times10^5$\msun\ by searching for broad \ha\ emission lines.
The AGN nature of 10 sources was further confirmed through the detection of X-ray emission,
defining a bona fide IMBH subsample.
Our catalog includes 12 of their sources in common with their candidate list,
four of which belong to their bona fide IMBH sample.
Most of the remaining sources are excluded from our analysis because,
under our uniform emission-line decomposition and statistical selection criteria,
the putative broad \ha\ component is either too weak or not statistically significant.
These differences in the adopted narrow broad-line decomposition strategies
and in the statistical criteria used to identify significant broad components
may explain the systematically lower BH mass estimates
in the \citet{chilingarian18} candidate sample.

\citet{Pucha2025} constructed a DESI-based sample 
that extends the redshift range for broad \ha\ 
to $z\approx0.45$ accessible by DESI's spectrograph.
They visually selected broad-line AGNs with \mbh$\leqslant10^{6}$\msun\ 
and classified 151 sources as confident and 147 tentative, 
yielding a total of 298 IMBH candidates,
which they highlighted as the largest sample to date.
Using an independent SDSS-based dataset and a quantitatively defined selection procedure,
our catalog contains 313 confident IMBH AGNs with \mbh$\leqslant10^{6}$\msun.
Because \citet{Pucha2025} did not release the coordinates (RAs and decl.) of their sources, 
a direct object-by-object comparison between the two samples is not possible.

\begin{figure*}[htbp]%
	\centering
	\includegraphics[width=0.9\textwidth]{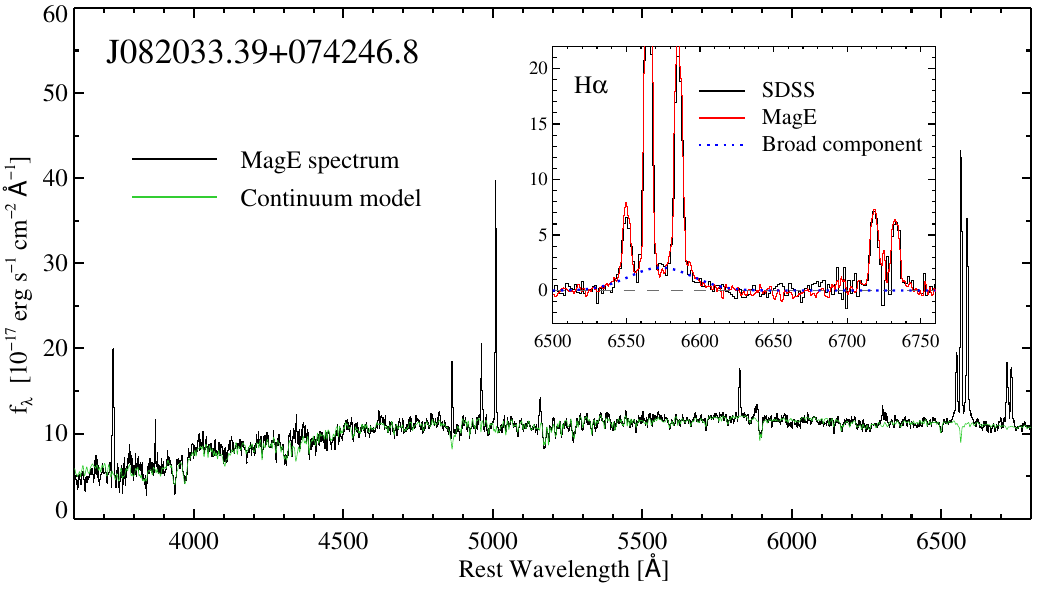}
        \includegraphics[width=0.9\textwidth]{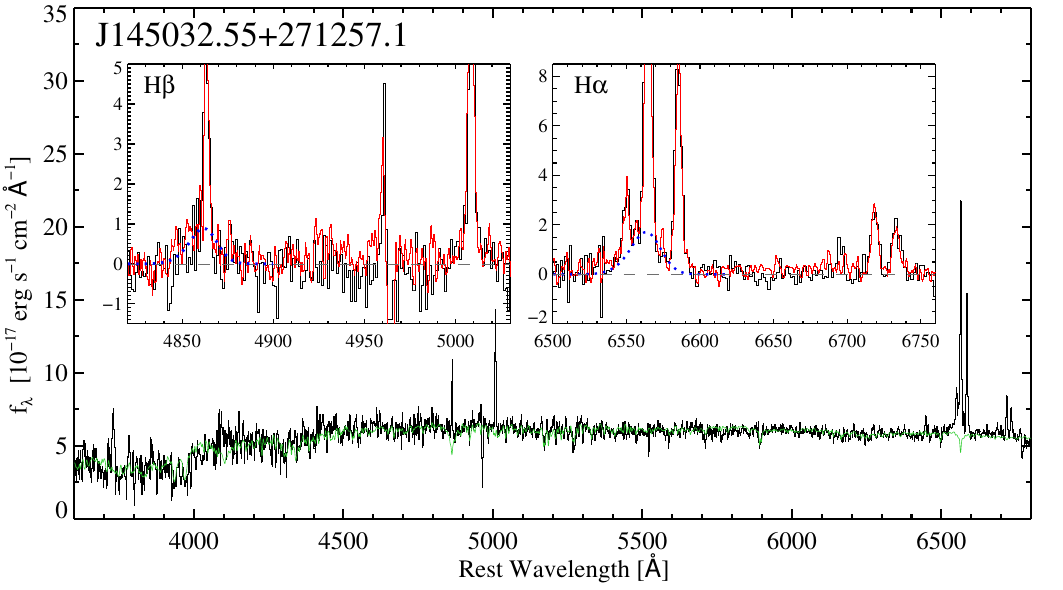}
	\caption{Illustration of the spectra of MagE (red) and SDSS (black) 
                for an IMBH AGN (J082033.39$+$074246.8) and an object classified as ``uncertain'' (J145032.55$+$271257.1) by our pipeline. 
                The green lines indicate the fitted continua for MagE and SDSS spectra.
                The inset panels provide a zoomed-in view of \hb\ and \ha\ regions in the 
                MagE and SDSS spectra (continuum subtracted), 
                with the MagE spectrum scaled to match the SDSS spectrum based on \oiii\,$\lambda5007$ flux.
                }
	\label{fig:fig3}
\end{figure*}

Figure~\ref{fig:fig2} compares our IMBH-AGN sample with those built in earlier studies 
\citep{gh07,dong12,liu18} in terms of the redshift distribution (panel~(a)),
the joint distribution of broad \ha\ flux versus FWHM (panel~(b)).
Because neither \citet{chilingarian18} nor \citet{Pucha2025} published measurements of broad-line FWHM or flux,
their samples could not be included in this comparison.
Compared with previous studies, our sample extends to higher redshifts.
In the log\,$f_\mathrm{\bha}$--log\,FWHM$_\mathrm{\bha}$ plane, 
which simultaneously reflects the strength and width of the broad-line emission,
our sources probe broad-\ha\ fluxes $\sim 0.4$ dex fainter than those in earlier samples.
The newly identified $z>0.35$ sources occupy a region characterized by 
both the weakest and narrowest broad-line features (see the red solid circles in the bottom-left corner of panel (b)), 
underscoring the improved sensitivity and refined fitting procedures of this work 
that enable the detection of such challenging sources in SDSS spectra.
In our fitting procedure, no hard requirement is imposed on the overall spectral S/N prior to analysis. 
This choice is motivated by the fact that some sources exhibit relatively low continuum quality 
(median spectral S/N $\leqslant$ 5), 
while their emission lines still process sufficient significance to allow detection of broad components,
as illustrated by the example shown in Figure~\ref{fig:fig1}, panel (c),
where a spectrum with median S/N $\approx$ 3.9 nevertheless displays a clearly identifiable broad \ha\ line.
As a result, the final IMBH sample spans a wide range of spectral quality, 
with median S/N (per pixel) values ranging from 2.8 to 77. 
The distribution peaks at a median S/N of $\approx$ 17, 
and 51 out of the 930 objects have median S/N below 5. 
This broad S/N distribution reflects our strategy of evaluating candidates 
through spectral modeling rather than preselecting sources based on continuum data quality alone.

We compare the virial BH masses derived in this work 
with those reported in the literature for overlapping objects. 
The agreement is good, with small systematic offsets:
0.07 dex (scatter 0.17 dex) relative to \citet{gh07},
and 0.01 dex (scatter 0.16 dex) for the combined \citet{dong12} and \citet{liu18} sample,
where $\Delta$\,log\,\mbh $=$ log\,\mbh\ (this work) $-$ log\,\mbh\ (literature).
These small offsets indicate no significant systematic difference in mass scale.
The remaining scatter is well within the typical uncertainties of single-epoch virial mass estimators 
and likely reflects differences in spectral decomposition procedures, 
particularly in the treatment of broad-narrow line separation.

\subsection{Supporting Observations}

To further validate the reliability of broad-line detections in our IMBH sample, 
especially for sources with low S/N in the SDSS spectra, 
we make use of two complementary spectroscopic datasets outside the SDSS campaign.

We first examine a set of observations conducted in 2017 March 24-25 and July 20-22 using the 
Magellan Echellette (MagE) Spectrograph \citep{mage} on the Magellan 6.5 m Baade telescope at Las Campanas Observatory.
These observations were carried out as a follow-up to the IMBH sample study of \citet{dong12}, 
which provides the foundation for the present study.
The MagE observations were conducted with two exposures per target, 
yielding a total integration time of 1800--3600\,s. 
The resulting spectra have a higher spectral resolution ($R \approx 4000$) compared to SDSS, 
enabling more reliable decomposition of broad and narrow line components, 
particularly in sources with complex or blended profiles. 
While the overall S/N of the MagE spectra in the \ha\ region is comparable to or slightly higher than that of the SDSS spectra, 
the improved resolution is the primary factor that allows more robust identification 
and characterization of broad components in previously ``uncertain'' sources.

The observed targets include 30 sources previously classified as ``uncertain'' due to 
marginal broad-line detections in the SDSS spectra by \citet{dong12},
with the goal of identifying AGNs powered by black holes with masses below $10^{5}$\msun.
Among them, 14 sources are confirmed to exhibit broad lines based on the MagE spectra,
including six that show significantly enhanced broad \ha\ compared to their earlier SDSS spectra (see \citealt{liu2021} for details).
The remaining 16 sources either show no significant broad components or 
remain ambiguous due to weak emission relative to noise and host galaxy contamination.

Figure~\ref{fig:fig3} presents the MagE spectra for two representative objects.
J0820$+$0742 was originally identified as ``uncertain'' by \citet{dong12}, 
but is re-identified as an IMBH in this work based on updated spectral decomposition.
In its SDSS spectrum, the identification of a broad \ha\ component was debatable, 
primarily due to the relatively broad \sii\ lines.
When \sii\ was used to constrain the narrow-line profile, the inferred broad \ha\ component appeared weak.
However, the MagE spectrum reveals a more extended broad \ha\ blue wing,
supporting the presence of a broad component in the decomposition. 
The object J1450$+$2712, classified as ``uncertain'' both in \citet{dong12} and this work, 
displays a clear, broad \hb\ profile in the MagE spectrum.
The higher spectral resolution of MagE resolves the stellar absorption features more clearly,
allowing for a more accurate fit and subtraction of the host galaxy continuum.

\begin{figure*}[htbp]%
	\centering
	\includegraphics[width=1\textwidth]{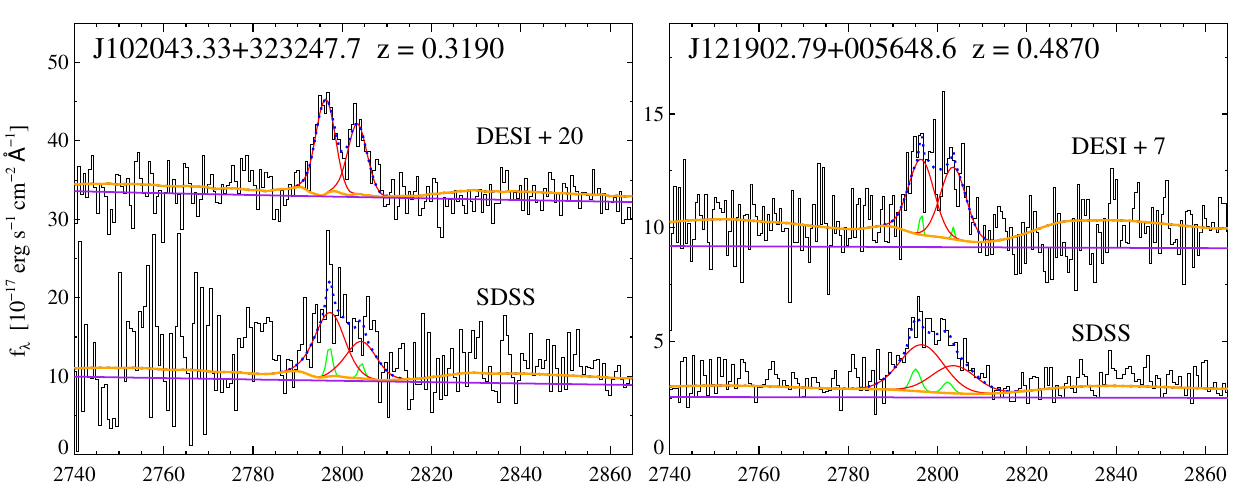}
    \includegraphics[width=1\textwidth]{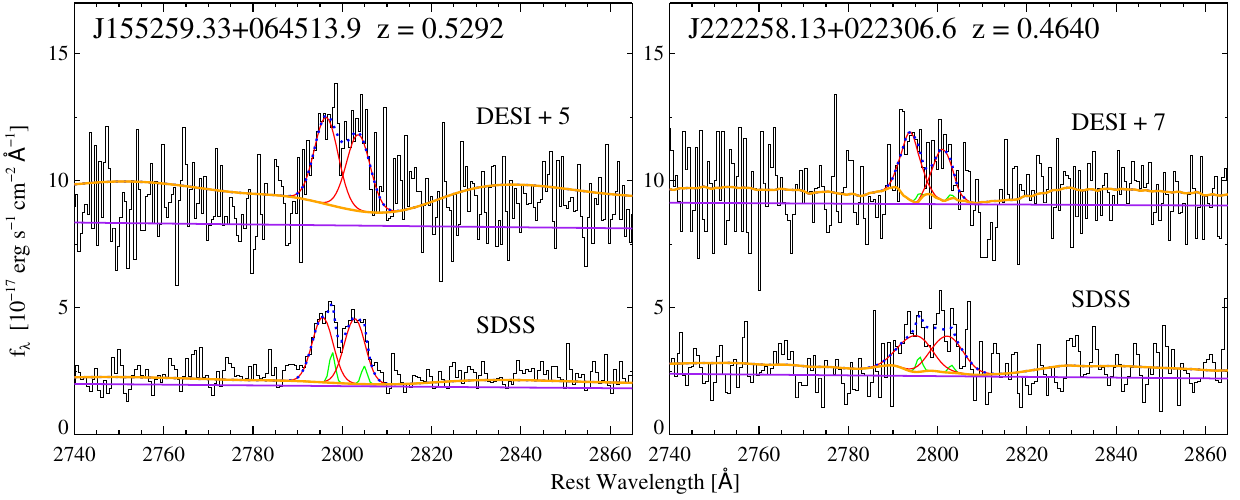}
	\caption{Illustration of \mgii\ emission-line fitting for four sources with 
             both SDSS and DESI spectra.
             Each panel focuses on the rest-frame wavelength range 2740--2860\,\AA, 
             covering the \mgii\,$\lambda\lambda2796,2803$ doublet. 
             Both SDSS and DESI spectra are plotted in black; 
             the DESI spectra are vertically offset for clarity, with the offset values labeled in the figure.   
             No spectral smoothing is applied.
             Colored lines represent the best-fit models for the emission-line and continuum components, 
             following the same color scheme as in Figure~\ref{fig:fig1}.}
	\label{fig:fig4}
\end{figure*}

In addition to the MagE spectra, 
we further assess the reliability of our broad \mgii\ identifications using newly available data from the 
Dark Energy Spectroscopic Instrument (DESI; \citealp{DESI2016}) Data Release 1 \citep[DR1][]{DESIDR1}.
DESI spectra span a wavelength range of 3600-9800\,\AA\ and are obtained using three spectrographs:
the blue (3600-5800\,\AA), red (5600-7600\,\AA), and near-infrared (7400-9800\,\AA).
The spectral resolving power increases from $R \sim 2000$ at 3600\,\AA\, to $R \sim 5500$ at 9800\,\AA\, 
offering approximately twice the resolution of the SDSS spectra.
We cross-match our SDSS IMBH sample with the DESI DR1 spectra using a 3\arcsec\ matching radius. 
After excluding spectra corresponding to neighboring objects, we finally get 327 matches in total. 
We select a subset with $z \geqslant 0.3$, 
where the \mgii\ line falls within the effective DESI wavelength coverage,
resulting 17 objects.
We apply the same \mgii\ fitting methodology used for the SDSS sample (\S~\ref{subsec:mg2fit})
and identify eight sources with clear broad \mgii\ emission lines.
Notably, all eight are included in the sample of 24 IMBHs with significant \mgii\ detections previously identified from SDSS, 
providing an independent spectroscopic confirmation.
Among the remaining nine sources, 
one object shows a tentative \mgii\ excess in the smoothed spectrum 
but is not included as a secure detection because the feature is not statistically robust.
The other sources do not yield convincing broad \mgii\ signatures in either DESI or SDSS data, 
owing to limited spectral quality in the \mgii\ region.

Figure~\ref{fig:fig4} presents four representative examples 
(J1020$+$3232, J1219$+$0056, J1552$+$0645, and J2222$+$0223),
each showing consistent broad \mgii\ features in both SDSS and DESI spectra.
For J1020$+$3232, the DESI spectrum has significantly higher S/N than the SDSS spectrum 
and offers a more clearly resolved \mgii\ profile. 
The measured FWHM are $838 \pm 150$\,\kms\ from SDSS and $543 \pm 14$\,\kms\ from DESI.
For J1219$+$0056 and J2222$+$0223, the DESI spectra have somewhat lower S/N compared to SDSS,
but still better resolve the \mgii\ doublet structures.
The corresponding FWHMs are:
$1231 \pm 168$\,\kms\ (SDSS) versus $717 \pm 58$\,\kms\ (DESI) for J1219+0056,
and $871\pm326$\,\kms\ (SDSS) versus $506 \pm 50$\,\kms\ (DESI) for J2222+0223.
These comparisons illustrate the advantage of DESI's improved spectral resolution, 
even when the S/N is lower.
For J1552$+$0645, the SDSS spectrum has significantly higher S/N 
and better separates the broad and narrow components of the \mgii\ doublet. 
While DESI also shows the same features, the narrow-line signature is less distinct.
The FWHM derived from SDSS and DESI are: $564 \pm 23$\,\kms\ (SDSS) versus $665 \pm45 $\,\kms\ (DESI).
Taken together, these examples demonstrate that, at comparable or even slightly lower spectral S/N levels,
DESI spectra offer better resolution of the \mgii\ line profiles than SDSS.
This suggests that DESI data may serve as a valuable resource for the identification of IMBHs at higher redshifts 
($z \approx 0.5-2$),
where traditional optical diagnostics such as \ha\ and \hb\ are redshifted out of the observed optical window.

\section{Results}

\begin{figure}[htbp]%
	\centering
	\includegraphics[width=0.495\textwidth]{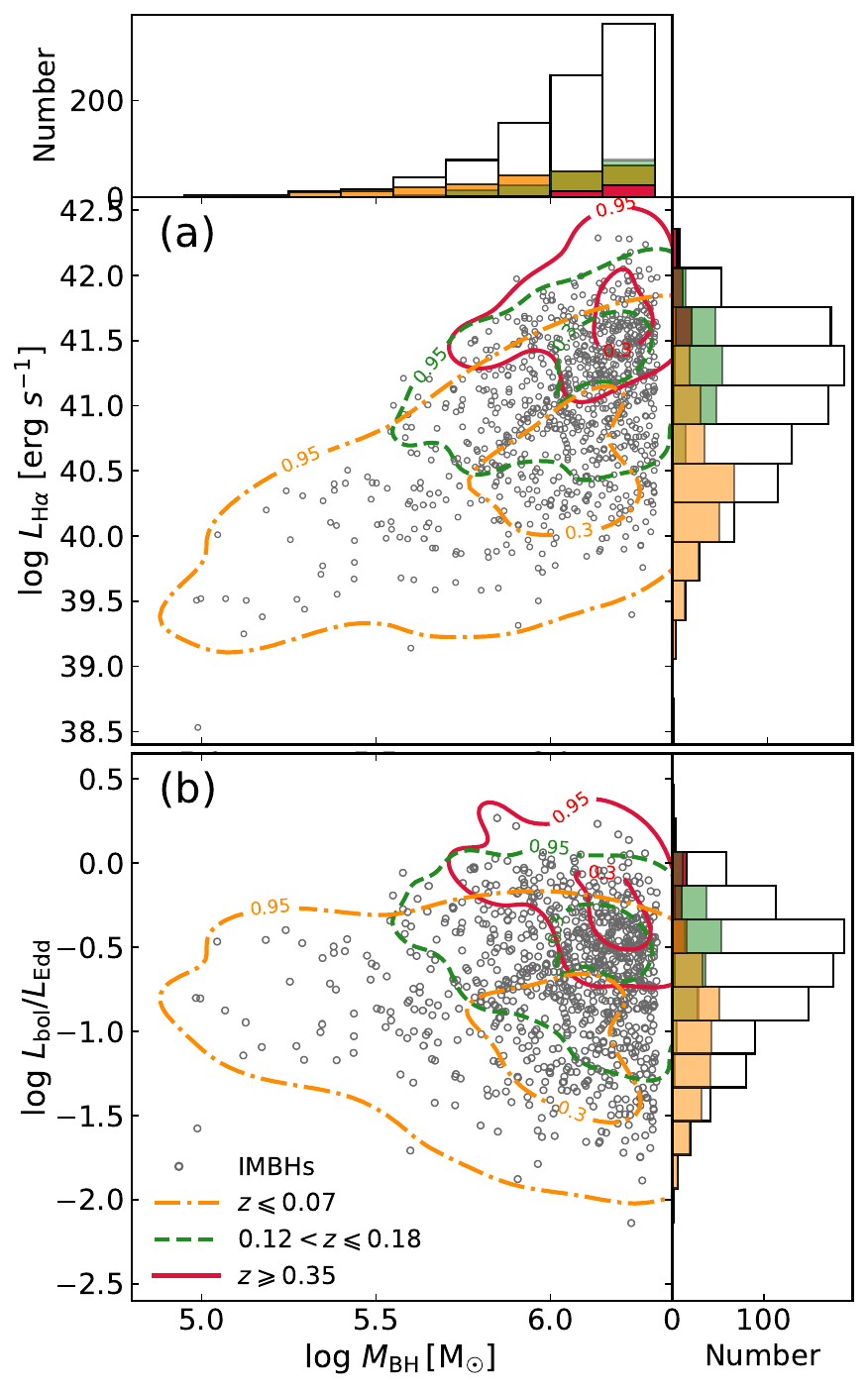}
	\caption{Distributions of our IMBH sample on the (a) log\,\mbh-log\,$L_{\ha}$ 
             and (b) log\,\mbh-log\,\redd\ planes. Gray open circles represent all IMBHs.
             The orange, green, and red contours indicate the subsamples within three redshift bins:
             $z\leqslant0.07$, $0.12<z<0.18$, and $z>0.35$, respectively.
             Each set of contours encloses 30\% and 95\% of the respective subsample.
             The histograms along the top and right margins show the distributions of BH mass, Eddington ratio, and broad \ha\ luminosity for the full sample (black),
             overlaid with the corresponding distributions for the three redshift subsamples in matching colors.}
	\label{fig:fig5}
\end{figure}

The final sample spans a redshift range of $z=0-0.57$, 
with a significant extension into the redshift range of $z=0.35-0.57$, 
as illustrated in Figure~\ref{fig:fig2}a.
Figure~\ref{fig:fig5} displays the distribution of the sample 
in the $\log\mbh-\log L_\mathrm{\bha}$ and $\log\mbh-\log\redd$ planes.
The broad \ha\ luminosities, $L_\mathrm{\bha}$, 
span from $10^{37.0}$ to $10^{42.3}$ \ergs, 
with a median of $10^{40.7}$ \ergs. 
The BH masses range from $1.0\times10^{4}$ to $2.0\times10^{6}$ \msun,
and Eddington ratios span from 0.01 to 1.9, with the median value of 0.23.

Here we present the basic distributions of several key physical quantities derived from spectral analysis of the sample, 
including BH mass, Eddington ratio, broad \ha\ luminosity, 
emission-line diagnostics, and a brief discussion of the \mgii\ line widths.
Possible trends with redshift are discussed in the following subsection.

\subsection{Narrow-line Diagnostic Diagrams}

\begin{figure*}[tbp]%
	\centering
	\includegraphics[width=\textwidth]{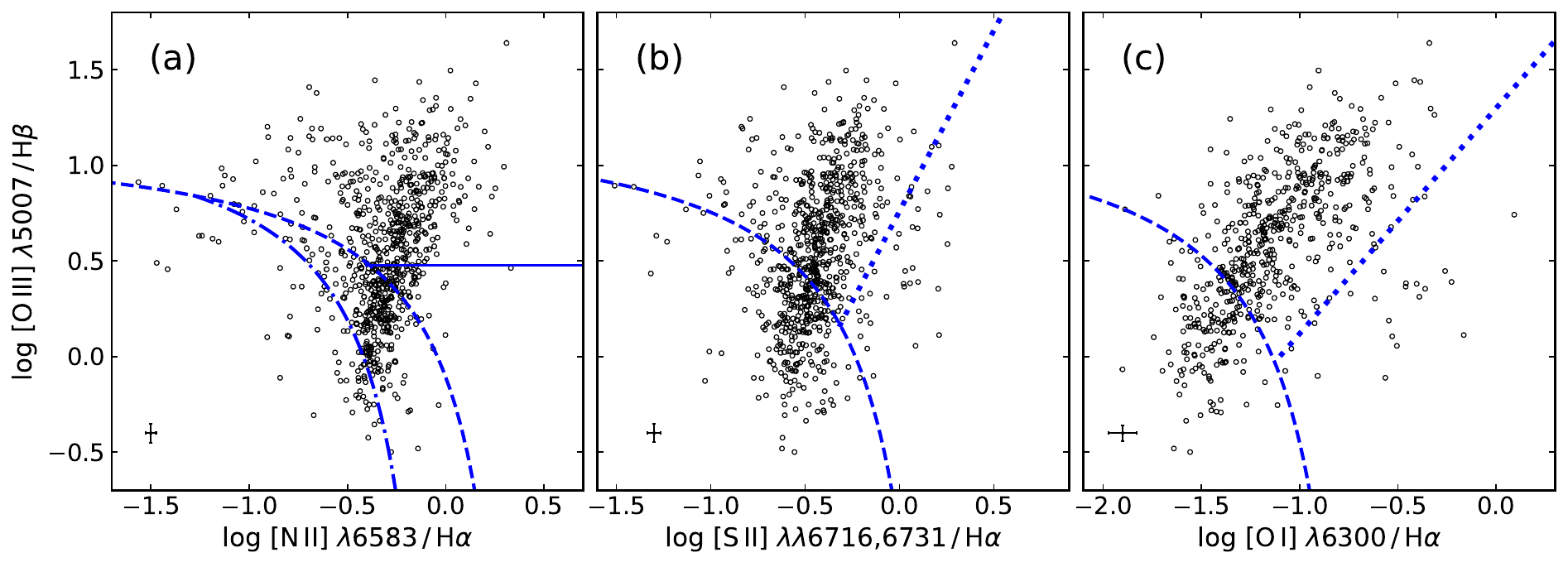}
	\caption{Narrow-line diagnostic diagrams of our IMBH sample, 
    showing \oiii\,$\lambda 5007$/\hb\ vs. (a) \nii\,$\lambda 6583$/\ha, (b) \sii\,$\lambda\lambda6716,6731$/\ha, and (c) \oi\,$\lambda 6300$/\ha.
    The dashed and dotted blue curves separating \hii\ regions, Seyfert galaxies, and LINERs are taken from \citet{kewley01} and \citet{kewley06}, respectively.
    In panel (a), the dash-dotted line shows the empirical division between star-forming galaxies and AGNs from \citet{Kauffmann03a}, and the solid horizontal line marks \oiii\,$\lambda5007$/\hb\ = 3,  traditionally used to distinguish Seyfert galaxies and LINERs. 
    A typical 1 $\sigma$ error bar is shown in each panel.}
	\label{fig:fig6}
\end{figure*}

Specific narrow emission-line ratios provide valuable diagnostics for identifying 
the dominant ionizing sources that excite the circumnuclear gas in galaxies.
The diagrams, incorporating narrow-line ratios such as \oiii\,$\lambda5007$/\hb, \nii\,$\lambda6583$/\ha, \sii\,$\lambda\lambda6716,6731$/\ha, and \oi\,$\lambda6300$/\ha,
have been widely used to distinguish between Seyfert galaxies, 
low-ionization nuclear emission-line region sources (LINERs; \citealt{heckman1980}), 
and star-forming galaxies \citep{BPT1981,Veilleux&Osterbrock1987,Ho1997a,kewley01,kewley06,kauffmann03b}.
Figure~\ref{fig:fig6} displays the distribution of our IMBH AGNs across these diagrams.
Overall, the distribution of our sample across the narrow-line diagnostic diagrams 
is broadly consistent with previous IMBH studies \citep{gh07,dong12,liu18},
spanning the expected range of line ratios for such systems.
In the \oiii\,$\lambda5007$/\hb\ vs. \nii\,$\lambda6583$/\ha\ diagram,
57\% of the sources fall within the Seyfert region, 35\% lie in the composite zone,
and only 8\% are classified as pure star-forming galaxies.
In the \oiii\,$\lambda5007$/\hb\ versus \sii\,$\lambda\lambda6716,6731$/\ha\ and 
\oi\,$\lambda6300$/\ha\ diagrams, 
approximately two-thirds of the objects occupy the AGN region,
with the remaining located in the \hii\ region.
Roughly 30 objects in our sample show LINER-like characteristics,
primarily based on the Seyfert-LINER demarcation lines from \citet{kewley06} in panels (b) and (c).
Visual inspection of the SDSS spectra for these 30 LINER-like IMBHs reveals that nearly all 
show features of old stellar populations.
Similar trends have been noted for low-mass AGNs residing in red-sequence hosts \citep[e.g.,][]{liu18}, 
suggesting that LINER-like spectra in IMBH systems are likely associated with evolved host galaxies,
consistent with the general tendency of LINERs to reside in galaxies with older stellar populations \citep{Ho1997c}.
Examples include J0838+5406 \citep{j0838} and J1342+0223 \citep{liu2021}, 
whose spectra show red continua dominated by older stellar populations.
The Eddington ratios for these LINERs span from 0.007 to 0.43, 
with a median value of 0.06.
This range is broadly consistent with the picture that LINERs generally have low accretion rates \citep[e.g.,][]{Ho2008,Ho2009}.

\begin{figure}[tbp]%
	\centering
        \includegraphics[width=0.495\textwidth]{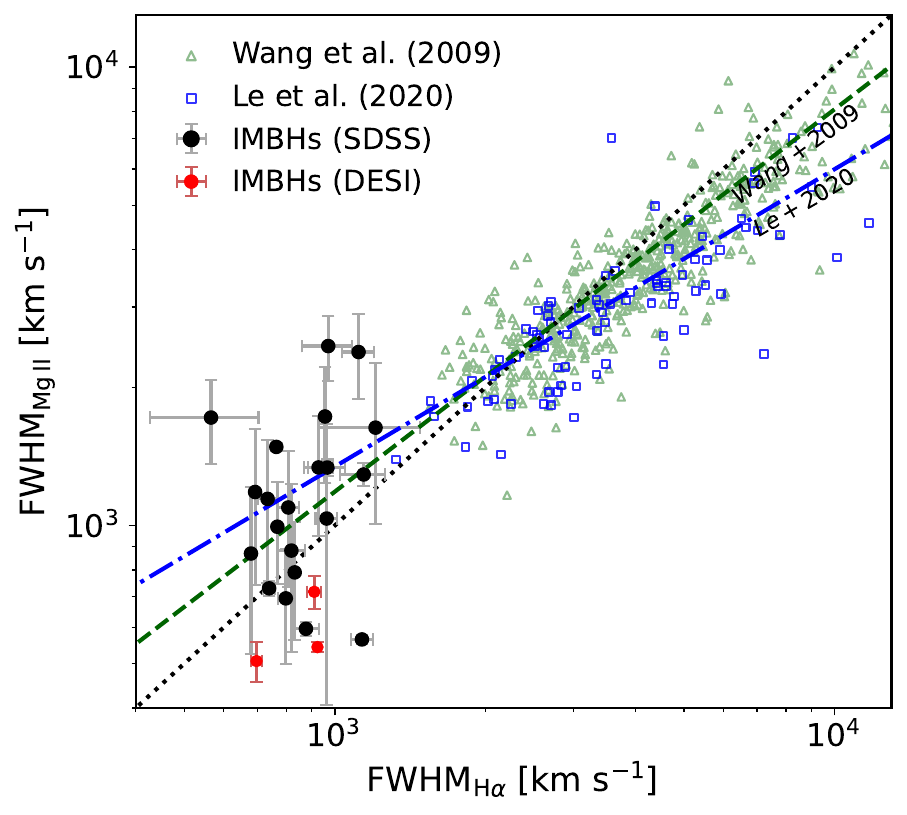}
	\caption{Comparison of broad-line widths: 
             FWHM$_{\mathrm{Mg}\,\scriptscriptstyle\mathrm{II}}$ vs FWHM$_{\ha}$ 
             for 24 IMBHs with detectable \mgii\ emission (black and red dots). 
             Black points represent measurements from SDSS spectra. 
             Red points highlight three sources (J1020$+$3232, J1219$+$0056, and J2222$+$0223)
             for which we show DESI DR1 measurements for display because the DESI-based fits 
             are of higher quality than the SDSS-based fits (their DESI \mgii\ fits are shown in Figure~\ref{fig:fig4}).
             Error bars indicate measurement uncertainties.
             Green triangles are from \citet{wang09},
             and blue squares represent the combined sample from \citet{Woo18} and \citet{Bark19}, 
             which are used in \citet{Le&Woo2020}.
             FWHM$_{\hb}$ values from these reference samples are converted to FWHM$_{\ha}$ 
             using the empirical relation from \citet{gh05}. 
             The black dotted line marks the 1:1 relation;
             blue dashed and green dash-dotted lines show the 
             FWHM$_{\mathrm{Mg}\,\scriptscriptstyle\mathrm{II}}$-FWHM$_{\hb}$ relations from \citet{wang09} and \citet{Le&Woo2020}, 
             respectively, with FWHM$_{\hb}$ converted to FWHM$_{\ha}$.
                 }
    \label{fig:fig7}
\end{figure}

\subsection{Broad \mgii\ Emission in IMBHs}

Apart from a few case studies \citep[e.g., NGC 4395;][]{NGC4395_Filippenko93,NGC4395_Kraemer99}
our knowledge of UV emission lines in IMBHs has long been limited 
by the lack of suitable spectroscopic data.
Using the extended wavelength range of BOSS/eBOSS and our \mgii\ fitting approach,
we identify 24 sources with reliable broad \mgii\ components.
Among them, eight sources are independently confirmed by higher-resolution DESI spectra (Figure~\ref{fig:fig4}), 
further validating the robustness of the detections.
The measured FWHM of broad \mgii, after correcting for instrumental broadening, 
spans from 565 to 2460 \kms.
This range is broadly consistent with the low FWHM extension of the empirical \mgii-\ha\ line width relation established for more massive AGNs \citep[e.g.,][see Figure~\ref{fig:fig7}]{wang09,Le&Woo2020}.

The distribution of FWHM for broad \mgii\ and \ha\ for the 24 identified sources 
shows significant scatter (Figure~\ref{fig:fig7}).
For three of them (J1020$+$3232, J1219$+$0056, and J2222$+$0223), 
the DESI spectra simultaneously cover both \mgii\ and \ha\ regions,
and their DESI-based fits of both lines are of higher quality than the corresponding SDSS spectra.
For display purposes,
we therefore adopt the DESI-based measurements for these three sources in the figure 
(marked as red dots; see also Figure~\ref{fig:fig4} and \S~\ref{subsec:mg2fit} for details).
For comparison, we overlay the FWHM$_{\ha}$-FWHM$_{\mathrm{Mg}\,\scriptscriptstyle\mathrm{II}}$ 
distributions for luminous, more massive AGNs from \citet{wang09} and \citet{Le&Woo2020}, 
whose compilation derives from \citet{Woo18} and \citet{Bark19}.
Since these reference studies report FWHM of broad \hb, 
we convert their FWHM$_{\hb}$ values to FWHM$_{\ha}$ using the empirical relation from \citet{gh05}.
We also display the corresponding best-fit FWHM$_{\hb}$-FWHM$_{\mathrm{Mg}\,\scriptscriptstyle\mathrm{II}}$ relations from both studies,
with FWHM$_{\hb}$ again converted to FWHM$_{\ha}$ for consistency.
Our IMBHs extend the FWHM$_{\ha}$-FWHM$_{\mathrm{Mg}\,\scriptscriptstyle\mathrm{II}}$ distribution to the low-FWHM regime, 
occupying a region that has been largely inaccessible to previous calibrations based on more massive broad-line AGN sample.

On average, the measured \mgii\ widths appear slightly broader than those of \ha,
with more sources lying above the one-to-one line in the 
FWHM$_{\ha}$-FWHM$_{\mathrm{Mg}\,\scriptscriptstyle\mathrm{II}}$ plane,
This trend is largely consistent with the low-FWHM extrapolation of the empirical relations 
from \citet{wang09} and \citet{Le&Woo2020}, 
which are derived from higher-mass broad-line AGN samples.
However, the current \mgii\ line measurements in IMBH AGNs have substantial uncertainties, 
and extrapolations from different high-mass samples diverge significantly at the low-mass end, 
would leading to large potential differences in \mgii-based BH mass estimates.
At present, no direct calibration exists for the FWHM$_{\ha}$-FWHM$_{\mathrm{Mg}\,\scriptscriptstyle\mathrm{II}}$
relation in the low-mass regime based on IMBH data.
High-quality spectra covering both \ha\ (or \hb) and \mgii\ in low-mass AGNs are therefore essential 
to establish a robust relation. 
Improved calibration will enable more accurate \mgii-based BH mass estimates 
and facilitate the reliable identification of IMBHs at higher redshifts, 
where \ha\ and \hb\ lines are redshifted out of the optical window.
In this context, as illustrated by our comparison of DESI and SDSS fits (Figure~\ref{fig:fig4}), 
DESI spectra will offer a promising way for systematically searching for IMBHs at higher redshift.

\subsection{Probable Evidence for Ongoing Evolution of IMBHs AGNs at Low Redshift}

\begin{figure}[htbp]%
	\centering
	\includegraphics[width=0.494\textwidth]{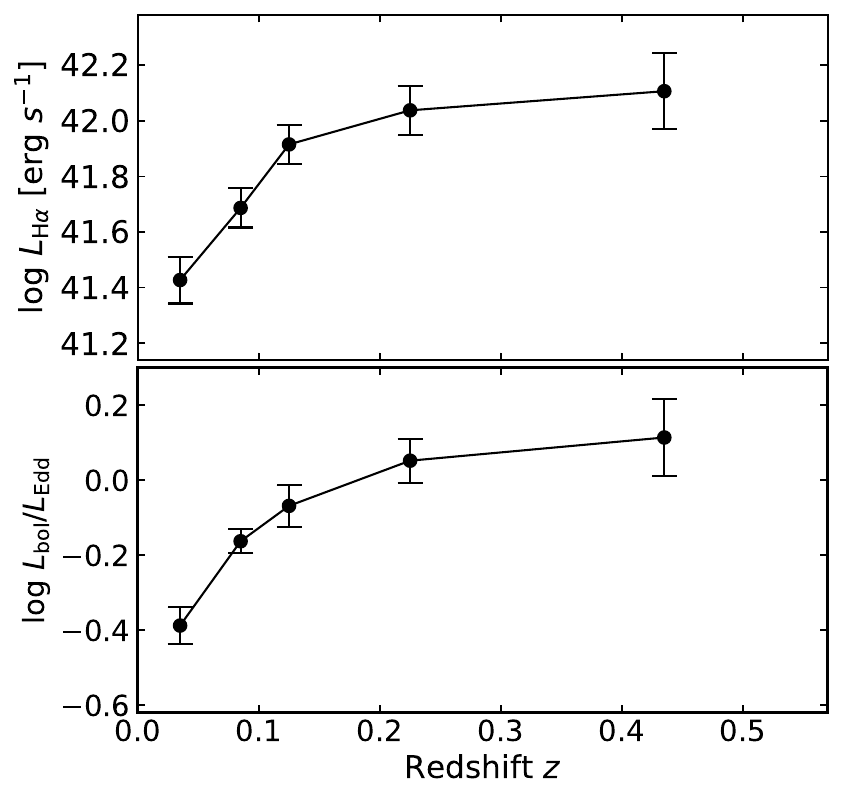}
	\caption{Mean value of top 10 sources with the hightest broad \ha\ luminosity 
    (top panel) or Eddington ratios (bottom panel) for five redshift intervals:
     $z=0-0.07$, $z=0.07-0.12$, $z=0.12-0.18$, $z=0.18-0.35$, and $z=0.35-0.57$.
     The corresponding standard deviations are indicated by error bars.} 
	\label{fig:fig8}
\end{figure}

The large size and extended redshift coverage ($z\approx0.6$) of our IMBH sample enable us
to investigate the cosmic evolution of their accretion properties at low redshift.
In Figure~\ref{fig:fig5}, 
we observe a trend that IMBHs occupy different regions in both
the log\mbh--log$L_{\ha}$ and log\mbh--log\redd\ planes, depending on redshift.
We focus on trends in directly observable quantities, using the broad \ha\ luminosity. 
The Eddington ratio, by contrast, is a derived quantity that is intrinsically tied to luminosity.
Notably, the upper envelope of the broad \ha\ luminosity distribution shows a pronounced decline 
toward lower redshift, 
and the corresponding behavior in Eddington ratio is presented as supporting information.
While such evolution is well known for high-mass AGNs (i.e., quasars), 
analogous evidence for low-mass AGNs has not been previously reported.
 
To characterize this behavior, we divide the sample into five redshift bins: 
$z=0-0.07$, $0.07-0.12$, $0.12-0.18$, $0.18-0.35$, and $0.35-0.57$.
These bins are chosen to balance statistical significance with temporal resolution.
The highest-redshift bin is set broader than the others to ensure sufficient sample size (47 objects),
while the remaining four bins each contain at least 150 sources.
Given that any observed distribution suffers various selection effects,
it is inappropriate to directly compare the average luminosities or Eddington ratios of AGNs across different redshift bins.
However, it remains meaningful to compare the maximum values of these quantities 
as a proxy for the upper envelope of the population at each epoch.
Specifically, we select the 10 sources with the highest broad \ha\ luminosities 
in each redshift interval and use their mean and standard deviation to characterize 
the maximum luminosity envelope and its uncertainty; 
the same procedure is applied to the Eddington ratios for comparison.

As shown in Figure~\ref{fig:fig8}, 
the maximum broad \ha\ luminosities of IMBH systematically decline with decreasing redshift,
and a similar trend is seen for the maximum Eddington ratio.
Importantly, given the nature of flux-limited samples, 
selection effects generally favor higher completeness at low redshifts,
especially for the most luminous or highly accreting AGNs.
Thus, the observed decline cannot be readily explained by the Malmquist bias,
which, if anything, would act to diminish rather than enhance such a trend.
In fact, such a cosmic decline has been pointed out in passing by \cite{dong12}  
for the entire broad-line AGN sample at $z<0.35$ selected from SDSS DR4.
This trend aligns well with the conventional paradigm of 
``cosmic downsizing'' \citep{cowie+96} of high-mass AGNs \citep[e.g.,][]{Ueda2003,Hasinger2005,shen&kelly2012} 
since cosmic noon ($z\approx2.5$).
Furthermore, several previous studies \citep{heckman04, gh07MF,Cho&WooMF2024} 
have reported that low-mass AGNs become the dominant AGN population at $z\lesssim 0.35$
in terms of comoving number density.
A A full characterization of the cosmic evolution of low-mass AGNs will 
ultimately require determination of the AGN luminosity function, BH mass function, 
and Eddington-ratio distribution function with careful correction for selection effects, 
which will be pursued in forthcoming work (Liu et al., in preparation).

\section{Summary}

IMBH AGNs provide crucial insights into the low-mass end of the nuclear BH population,
offering important observational constraints on BH growth and AGN activity in low-mass galaxies.
Large-scale optical spectroscopic surveys remain the most effective way for conducting 
a systematic census of AGNs across a wide dynamic range in BH mass and accretion rate.
In this study, we construct a sample of 930 IMBHs from SDSS DR17 spectra,
following the selection procedures developed by \citet{dong12} and \citet{liu19}, 
with several necessary improvements to enhance detection robustness and sensitivity.
Notably, our sample extends the redshift coverage of low-$z$ IMBHs from 
$z\lesssim0.35$ to $z\lesssim0.57$, 
drawing from a parent sample with over 45,000 broad-line AGNs.
A detailed analysis of the full broad-line AGN sample will be presented in a forthcoming paper.

We summarize the main findings of this study as follows:

(1) The BH masses of the sample span from $10^{4.0}$ to $10^{6.3}$ \msun,
with Eddington ratios ranging from 0.01 to 1.9 (median $\approx$ 0.23).
The broad \ha\ luminosities range from $10^{37.0}$ to $10^{42.3}$ \ergs.
These measurements establish one of the largest uniformly selected IMBH-AGN samples over a broad range of accretion properties, 
enabling statistical studies of low-mass BH activity across an extended redshift range.

(2) Among the sources with $z>0.3$, 
we identify 24 objects with detected broad \mgii\,$\lambda\lambda2796,2803$ emission.
Of these, 8 are independently confirmed by DESI DR1 spectra,
which provide higher-resolution observations and consistent broad \mgii\ detections.
This provides the first systematic detection of broad \mgii\ lines in IMBHs,
supporting the use of \mgii-based diagnostics to extend IMBH searches to higher redshifts
where traditional Balmer-line measurements become inaccessible.

(3) We observed a redshift-dependent decline in the upper envelopes of the Eddington ratio 
and broad \ha\ luminosity distributions.
While selection incompleteness may affect the detailed population statistics, 
it is unlikely to fully account for the systematic shift of the upper envelope with redshift. 
This trend may indicate a decrease in the typical accretion activity of IMBH AGNs toward lower redshift; 
a definitive assessment will rely on a quantitative determination of the AGN luminosity function, 
BH mass function, and Eddington ratio distribution function, 
with robust correction for selection effects, which we will pursue in forthcoming work based on this sample.



\section{Acknowledgments}

This work is supported by the National Natural Science Foundation of China (12373013) and the Yunnan Provincial Basic Research Program (202501AT070029).
L.C.H. was supported by the China Manned Space Program (CMS-CSST-2025-A09), the National Science Foundation of China (12233001).
S.Y. acknowledges the National Key R\&D Program of China (2025YFA1614101).
This work is based on observations obtained by the SDSS; 
we acknowledge the entire SDSS team for providing the data that made this work possible.
We also made use of spectra obtained with the Magellan Baade Telescope/MagE.

\bibliography{imbh_sample}{}

@ARTICLE{SDSSDR7,
       author = {{Abazajian}, Kevork N. and {Adelman-McCarthy}, Jennifer K. and {Ag{\"u}eros}, Marcel A. and {Allam}, Sahar S. and {Allende Prieto}, Carlos and {An}, Deokkeun and {Anderson}, Kurt S.~J. and {Anderson}, Scott F. and {Annis}, James and {Bahcall}, Neta A. and et al.},
        title = "{The Seventh Data Release of the Sloan Digital Sky Survey}",
      journal = {\apjs},
         year = 2009,
        month = jun,
       volume = {182},
       number = {2},
        pages = {543-558},
          doi = {10.1088/0067-0049/182/2/543},
archivePrefix = {arXiv},
       eprint = {0812.0649},
 primaryClass = {astro-ph},
       adsurl = {https://ui.adsabs.harvard.edu/abs/2009ApJS..182..543A}
}

@ARTICLE{SDSSDR17,
       author = {{Abdurro'uf} and {Accetta}, Katherine and {Aerts}, Conny and {Silva Aguirre}, V{\'\i}ctor and {Ahumada}, Romina and {Ajgaonkar}, Nikhil and {Filiz Ak}, N. and {Alam}, Shadab and {Allende Prieto}, Carlos and {Almeida}, Andr{\'e}s and et al.},
        title = "{The Seventeenth Data Release of the Sloan Digital Sky Surveys: Complete Release of MaNGA, MaStar, and APOGEE-2 Data}",
      journal = {\apjs},
         year = 2022,
        month = apr,
       volume = {259},
       number = {2},
          eid = {35},
        pages = {35},
          doi = {10.3847/1538-4365/ac4414},
archivePrefix = {arXiv},
       eprint = {2112.02026},
 primaryClass = {astro-ph.GA},
       adsurl = {https://ui.adsabs.harvard.edu/abs/2022ApJS..259...35A}
}

@ARTICLE{Bark19,
       author = {{Bahk}, Hyeonguk and {Woo}, Jong-Hak and {Park}, Daeseong},
        title = "{Calibrating Mg II-based Black Hole Mass Estimators with H{\ensuremath{\beta}} Reverberation Measurements}",
      journal = {\apj},
         year = 2019,
        month = apr,
       volume = {875},
       number = {1},
          eid = {50},
        pages = {50},
          doi = {10.3847/1538-4357/ab100d},
       adsurl = {https://ui.adsabs.harvard.edu/abs/2019ApJ...875...50B}
}

@ARTICLE{Barth2004,
       author = {{Barth}, Aaron J. and {Ho}, Luis C. and {Rutledge}, Robert E. and {Sargent}, Wallace L.~W.},
        title = "{POX 52: A Dwarf Seyfert 1 Galaxy with an Intermediate-Mass Black Hole}",
      journal = {\apj},
         year = 2004,
        month = may,
       volume = {607},
       number = {1},
        pages = {90-102},
          doi = {10.1086/383302},
archivePrefix = {arXiv},
       eprint = {astro-ph/0402110},
 primaryClass = {astro-ph},
       adsurl = {https://ui.adsabs.harvard.edu/abs/2004ApJ...607...90B}
}

@ARTICLE{bentz09,
       author = {{Bentz}, Misty C. and {Peterson}, Bradley M. and {Netzer}, Hagai and {Pogge}, Richard W. and {Vestergaard}, Marianne},
        title = "{The Radius-Luminosity Relationship for Active Galactic Nuclei: The Effect of Host-Galaxy Starlight on Luminosity Measurements. II. The Full Sample of Reverberation-Mapped AGNs}",
      journal = {\apj},
         year = 2009,
        month = may,
       volume = {697},
       number = {1},
        pages = {160-181},
          doi = {10.1088/0004-637X/697/1/160},
archivePrefix = {arXiv},
       eprint = {0812.2283},
 primaryClass = {astro-ph},
       adsurl = {https://ui.adsabs.harvard.edu/abs/2009ApJ...697..160B}
}

@ARTICLE{highz-agn,
       author = {{Bogd{\'a}n}, {\'A}kos and {Goulding}, Andy D. and {Natarajan}, Priyamvada and {Kov{\'a}cs}, Orsolya E. and {Tremblay}, Grant R. and {Chadayammuri}, Urmila and {Volonteri}, Marta and {Kraft}, Ralph P. and {Forman}, William R. and {Jones}, Christine and {Churazov}, Eugene and {Zhuravleva}, Irina},
        title = "{Evidence for heavy-seed origin of early supermassive black holes from a z {\ensuremath{\approx}} 10 X-ray quasar}",
      journal = {Nature Astronomy},
         year = 2024,
        month = jan,
       volume = {8},
        pages = {126-133},
          doi = {10.1038/s41550-023-02111-9},
archivePrefix = {arXiv},
       eprint = {2305.15458},
 primaryClass = {astro-ph.GA},
       adsurl = {https://ui.adsabs.harvard.edu/abs/2024NatAs...8..126B}
}

@ARTICLE{BPT1981,
       author = {{Baldwin}, J.~A. and {Phillips}, M.~M. and {Terlevich}, R.},
        title = "{Classification parameters for the emission-line spectra of extragalactic objects.}",
      journal = {\pasp},
         year = 1981,
        month = feb,
       volume = {93},
        pages = {5-19},
          doi = {10.1086/130766},
       adsurl = {https://ui.adsabs.harvard.edu/abs/1981PASP...93....5B}
}

@ARTICLE{BC03,
       author = {{Bruzual}, G. and {Charlot}, S.},
        title = "{Stellar population synthesis at the resolution of 2003}",
      journal = {\mnras},
         year = 2003,
        month = oct,
       volume = {344},
       number = {4},
        pages = {1000-1028},
          doi = {10.1046/j.1365-8711.2003.06897.x},
archivePrefix = {arXiv},
       eprint = {astro-ph/0309134},
 primaryClass = {astro-ph},
       adsurl = {https://ui.adsabs.harvard.edu/abs/2003MNRAS.344.1000B}
}

@ARTICLE{Cho&WooMF2024,
       author = {{Cho}, Hojin and {Woo}, Jong-Hak},
        title = "{Constraining the Low-mass End of the Black Hole Mass Function and the Active Fraction of the Intermediate-mass Black Holes}",
      journal = {\apj},
         year = 2024,
        month = jul,
       volume = {969},
       number = {2},
          eid = {93},
        pages = {93},
          doi = {10.3847/1538-4357/ad4966},
archivePrefix = {arXiv},
       eprint = {2405.09441},
 primaryClass = {astro-ph.GA},
       adsurl = {https://ui.adsabs.harvard.edu/abs/2024ApJ...969...93C}
}

@ARTICLE{chilingarian18,
       author = {{Chilingarian}, Igor V. and {Katkov}, Ivan Yu. and {Zolotukhin}, Ivan Yu. and {Grishin}, Kirill A. and {Beletsky}, Yuri and {Boutsia}, Konstantina and {Osip}, David J.},
        title = "{A Population of Bona Fide Intermediate-mass Black Holes Identified as Low-luminosity Active Galactic Nuclei}",
      journal = {\apj},
         year = 2018,
        month = aug,
       volume = {863},
       number = {1},
          eid = {1},
        pages = {1},
          doi = {10.3847/1538-4357/aad184},
archivePrefix = {arXiv},
       eprint = {1805.01467},
 primaryClass = {astro-ph.GA},
       adsurl = {https://ui.adsabs.harvard.edu/abs/2018ApJ...863....1C}
}

@ARTICLE{cowie+96,
       author = {{Cowie}, Lennox L. and {Songaila}, Antoinette and {Hu}, Esther M. and {Cohen}, J.~G.},
        title = "{New Insight on Galaxy Formation and Evolution From Keck Spectroscopy of the Hawaii Deep Fields}",
      journal = {\aj},
         year = 1996,
        month = sep,
       volume = {112},
        pages = {839},
          doi = {10.1086/118058},
archivePrefix = {arXiv},
       eprint = {astro-ph/9606079},
 primaryClass = {astro-ph},
       adsurl = {https://ui.adsabs.harvard.edu/abs/1996AJ....112..839C}
}

@ARTICLE{BOSS,
       author = {{Dawson}, Kyle S. and {Schlegel}, David J. and {Ahn}, Christopher P. and {Anderson}, Scott F. and {Aubourg}, {\'E}ric and {Bailey}, Stephen and {Barkhouser}, Robert H. and {Bautista}, Julian E. and {Beifiori}, Alessandra and {Berlind}, Andreas A. and {Bhardwaj}, Vaishali and {Bizyaev}, Dmitry and {Blake}, Cullen H. and {Blanton}, Michael R. and {Blomqvist}, Michael and {Bolton}, Adam S. and {Borde}, Arnaud and {Bovy}, Jo and {Brandt}, W.~N. and {Brewington}, Howard and {Brinkmann}, Jon and {Brown}, Peter J. and {Brownstein}, Joel R. and {Bundy}, Kevin and {Busca}, N.~G. and {Carithers}, William and {Carnero}, Aurelio R. and {Carr}, Michael A. and {Chen}, Yanmei and {Comparat}, Johan and {Connolly}, Natalia and {Cope}, Frances and {Croft}, Rupert A.~C. and {Cuesta}, Antonio J. and {da Costa}, Luiz N. and {Davenport}, James R.~A. and {Delubac}, Timoth{\'e}e and {de Putter}, Roland and {Dhital}, Saurav and {Ealet}, Anne and {Ebelke}, Garrett L. and {Eisenstein}, Daniel J. and {Escoffier}, S. and {Fan}, Xiaohui and {Filiz Ak}, N. and {Finley}, Hayley and {Font-Ribera}, Andreu and {G{\'e}nova-Santos}, R. and {Gunn}, James E. and {Guo}, Hong and {Haggard}, Daryl and {Hall}, Patrick B. and {Hamilton}, Jean-Christophe and {Harris}, Ben and {Harris}, David W. and {Ho}, Shirley and {Hogg}, David W. and {Holder}, Diana and {Honscheid}, Klaus and {Huehnerhoff}, Joe and {Jordan}, Beatrice and {Jordan}, Wendell P. and {Kauffmann}, Guinevere and {Kazin}, Eyal A. and {Kirkby}, David and {Klaene}, Mark A. and {Kneib}, Jean-Paul and {Le Goff}, Jean-Marc and {Lee}, Khee-Gan and {Long}, Daniel C. and {Loomis}, Craig P. and {Lundgren}, Britt and {Lupton}, Robert H. and {Maia}, Marcio A.~G. and {Makler}, Martin and {Malanushenko}, Elena and {Malanushenko}, Viktor and {Mandelbaum}, Rachel and {Manera}, Marc and {Maraston}, Claudia and {Margala}, Daniel and {Masters}, Karen L. and {McBride}, Cameron K. and {McDonald}, Patrick and {McGreer}, Ian D. and {McMahon}, Richard G. and {Mena}, Olga and {Miralda-Escud{\'e}}, Jordi and {Montero-Dorta}, Antonio D. and {Montesano}, Francesco and {Muna}, Demitri and {Myers}, Adam D. and {Naugle}, Tracy and {Nichol}, Robert C. and {Noterdaeme}, Pasquier and {Nuza}, Sebasti{\'a}n E. and {Olmstead}, Matthew D. and {Oravetz}, Audrey and {Oravetz}, Daniel J. and {Owen}, Russell and {Padmanabhan}, Nikhil and {Palanque-Delabrouille}, Nathalie and {Pan}, Kaike and {Parejko}, John K. and {P{\^a}ris}, Isabelle and {Percival}, Will J. and {P{\'e}rez-Fournon}, Ismael and {P{\'e}rez-R{\`a}fols}, Ignasi and {Petitjean}, Patrick and {Pfaffenberger}, Robert and {Pforr}, Janine and {Pieri}, Matthew M. and {Prada}, Francisco and {Price-Whelan}, Adrian M. and {Raddick}, M. Jordan and {Rebolo}, Rafael and {Rich}, James and {Richards}, Gordon T. and {Rockosi}, Constance M. and {Roe}, Natalie A. and {Ross}, Ashley J. and {Ross}, Nicholas P. and {Rossi}, Graziano and {Rubi{\~n}o-Martin}, J.~A. and {Samushia}, Lado and {S{\'a}nchez}, Ariel G. and {Sayres}, Conor and {Schmidt}, Sarah J. and {Schneider}, Donald P. and {Sc{\'o}ccola}, C.~G. and {Seo}, Hee-Jong and {Shelden}, Alaina and {Sheldon}, Erin and {Shen}, Yue and {Shu}, Yiping and {Slosar}, An{\v{z}}e and {Smee}, Stephen A. and {Snedden}, Stephanie A. and {Stauffer}, Fritz and {Steele}, Oliver and {Strauss}, Michael A. and {Streblyanska}, Alina and {Suzuki}, Nao and {Swanson}, Molly E.~C. and {Tal}, Tomer and {Tanaka}, Masayuki and {Thomas}, Daniel and {Tinker}, Jeremy L. and {Tojeiro}, Rita and {Tremonti}, Christy A. and {Vargas Maga{\~n}a}, M. and {Verde}, Licia and {Viel}, Matteo and {Wake}, David A. and {Watson}, Mike and {Weaver}, Benjamin A. and {Weinberg}, David H. and {Weiner}, Benjamin J. and {West}, Andrew A. and {White}, Martin and {Wood-Vasey}, W.~M. and {Yeche}, Christophe and {Zehavi}, Idit and {Zhao}, Gong-Bo and {Zheng}, Zheng},
        title = "{The Baryon Oscillation Spectroscopic Survey of SDSS-III}",
      journal = {\aj},
         year = 2013,
        month = jan,
       volume = {145},
       number = {1},
          eid = {10},
        pages = {10},
          doi = {10.1088/0004-6256/145/1/10},
archivePrefix = {arXiv},
       eprint = {1208.0022},
 primaryClass = {astro-ph.CO},
       adsurl = {https://ui.adsabs.harvard.edu/abs/2013AJ....145...10D}
}

@ARTICLE{eBOSS,
       author = {{Dawson}, Kyle S. and {Kneib}, Jean-Paul and {Percival}, Will J. and {Alam}, Shadab and {Albareti}, Franco D. and {Anderson}, Scott F. and {Armengaud}, Eric and {Aubourg}, {\'E}ric and {Bailey}, Stephen and {Bautista}, Julian E. and {Berlind}, Andreas A. and {Bershady}, Matthew A. and {Beutler}, Florian and {Bizyaev}, Dmitry and {Blanton}, Michael R. and {Blomqvist}, Michael and {Bolton}, Adam S. and {Bovy}, Jo and {Brandt}, W.~N. and {Brinkmann}, Jon and {Brownstein}, Joel R. and {Burtin}, Etienne and {Busca}, N.~G. and {Cai}, Zheng and {Chuang}, Chia-Hsun and {Clerc}, Nicolas and {Comparat}, Johan and {Cope}, Frances and {Croft}, Rupert A.~C. and {Cruz-Gonzalez}, Irene and {da Costa}, Luiz N. and {Cousinou}, Marie-Claude and {Darling}, Jeremy and {de la Macorra}, Axel and {de la Torre}, Sylvain and {Delubac}, Timoth{\'e}e and {du Mas des Bourboux}, H{\'e}lion and {Dwelly}, Tom and {Ealet}, Anne and {Eisenstein}, Daniel J. and {Eracleous}, Michael and {Escoffier}, S. and {Fan}, Xiaohui and {Finoguenov}, Alexis and {Font-Ribera}, Andreu and {Frinchaboy}, Peter and {Gaulme}, Patrick and {Georgakakis}, Antonis and {Green}, Paul and {Guo}, Hong and {Guy}, Julien and {Ho}, Shirley and {Holder}, Diana and {Huehnerhoff}, Joe and {Hutchinson}, Timothy and {Jing}, Yipeng and {Jullo}, Eric and {Kamble}, Vikrant and {Kinemuchi}, Karen and {Kirkby}, David and {Kitaura}, Francisco-Shu and {Klaene}, Mark A. and {Laher}, Russ R. and {Lang}, Dustin and {Laurent}, Pierre and {Le Goff}, Jean-Marc and {Li}, Cheng and {Liang}, Yu and {Lima}, Marcos and {Lin}, Qiufan and {Lin}, Weipeng and {Lin}, Yen-Ting and {Long}, Daniel C. and {Lundgren}, Britt and {MacDonald}, Nicholas and {Geimba Maia}, Marcio Antonio and {Malanushenko}, Elena and {Malanushenko}, Viktor and {Mariappan}, Vivek and {McBride}, Cameron K. and {McGreer}, Ian D. and {M{\'e}nard}, Brice and {Merloni}, Andrea and {Meza}, Andres and {Montero-Dorta}, Antonio D. and {Muna}, Demitri and {Myers}, Adam D. and {Nandra}, Kirpal and {Naugle}, Tracy and {Newman}, Jeffrey A. and {Noterdaeme}, Pasquier and {Nugent}, Peter and {Ogando}, Ricardo and {Olmstead}, Matthew D. and {Oravetz}, Audrey and {Oravetz}, Daniel J. and {Padmanabhan}, Nikhil and {Palanque-Delabrouille}, Nathalie and {Pan}, Kaike and {Parejko}, John K. and {P{\^a}ris}, Isabelle and {Peacock}, John A. and {Petitjean}, Patrick and {Pieri}, Matthew M. and {Pisani}, Alice and {Prada}, Francisco and {Prakash}, Abhishek and {Raichoor}, Anand and {Reid}, Beth and {Rich}, James and {Ridl}, Jethro and {Rodriguez-Torres}, Sergio and {Carnero Rosell}, Aurelio and {Ross}, Ashley J. and {Rossi}, Graziano and {Ruan}, John and {Salvato}, Mara and {Sayres}, Conor and {Schneider}, Donald P. and {Schlegel}, David J. and {Seljak}, Uros and {Seo}, Hee-Jong and {Sesar}, Branimir and {Shandera}, Sarah and {Shu}, Yiping and {Slosar}, An{\v{z}}e and {Sobreira}, Flavia and {Streblyanska}, Alina and {Suzuki}, Nao and {Taylor}, Donna and {Tao}, Charling and {Tinker}, Jeremy L. and {Tojeiro}, Rita and {Vargas-Maga{\~n}a}, Mariana and {Wang}, Yuting and {Weaver}, Benjamin A. and {Weinberg}, David H. and {White}, Martin and {Wood-Vasey}, W.~M. and {Yeche}, Christophe and {Zhai}, Zhongxu and {Zhao}, Cheng and {Zhao}, Gong-bo and {Zheng}, Zheng and {Ben Zhu}, Guangtun and {Zou}, Hu},
        title = "{The SDSS-IV Extended Baryon Oscillation Spectroscopic Survey: Overview and Early Data}",
      journal = {\aj},
         year = 2016,
        month = feb,
       volume = {151},
       number = {2},
          eid = {44},
        pages = {44},
          doi = {10.3847/0004-6256/151/2/44},
archivePrefix = {arXiv},
       eprint = {1508.04473},
 primaryClass = {astro-ph.CO},
       adsurl = {https://ui.adsabs.harvard.edu/abs/2016AJ....151...44D}
}

@ARTICLE{DESI2016,
       author = {{DESI Collaboration} and {Aghamousa}, Amir and {Aguilar}, Jessica and {Ahlen}, Steve and {Alam}, Shadab and {Allen}, Lori E. and {Allende Prieto}, Carlos and {Annis}, James and {Bailey}, Stephen and {Balland}, Christophe and {Ballester}, Otger and {Baltay}, Charles and {Beaufore}, Lucas and {Bebek}, Chris and {Beers}, Timothy C. and {Bell}, Eric F. and {Bernal}, Jos{\'e} Luis and {Besuner}, Robert and {Beutler}, Florian and {Blake}, Chris and {Bleuler}, Hannes and {Blomqvist}, Michael and {Blum}, Robert and {Bolton}, Adam S. and {Briceno}, Cesar and {Brooks}, David and {Brownstein}, Joel R. and {Buckley-Geer}, Elizabeth and {Burden}, Angela and {Burtin}, Etienne and {Busca}, Nicolas G. and {Cahn}, Robert N. and {Cai}, Yan-Chuan and {Cardiel-Sas}, Laia and {Carlberg}, Raymond G. and {Carton}, Pierre-Henri and {Casas}, Ricard and {Castander}, Francisco J. and {Cervantes-Cota}, Jorge L. and {Claybaugh}, Todd M. and {Close}, Madeline and {Coker}, Carl T. and {Cole}, Shaun and {Comparat}, Johan and {Cooper}, Andrew P. and {Cousinou}, M. -C. and {Crocce}, Martin and {Cuby}, Jean-Gabriel and {Cunningham}, Daniel P. and {Davis}, Tamara M. and {Dawson}, Kyle S. and {de la Macorra}, Axel and {De Vicente}, Juan and {Delubac}, Timoth{\'e}e and {Derwent}, Mark and {Dey}, Arjun and {Dhungana}, Govinda and {Ding}, Zhejie and {Doel}, Peter and {Duan}, Yutong T. and {Ealet}, Anne and {Edelstein}, Jerry and {Eftekharzadeh}, Sarah and {Eisenstein}, Daniel J. and {Elliott}, Ann and {Escoffier}, St{\'e}phanie and {Evatt}, Matthew and {Fagrelius}, Parker and {Fan}, Xiaohui and {Fanning}, Kevin and {Farahi}, Arya and {Farihi}, Jay and {Favole}, Ginevra and {Feng}, Yu and {Fernandez}, Enrique and {Findlay}, Joseph R. and {Finkbeiner}, Douglas P. and {Fitzpatrick}, Michael J. and {Flaugher}, Brenna and {Flender}, Samuel and {Font-Ribera}, Andreu and {Forero-Romero}, Jaime E. and {Fosalba}, Pablo and {Frenk}, Carlos S. and {Fumagalli}, Michele and {Gaensicke}, Boris T. and {Gallo}, Giuseppe and {Garcia-Bellido}, Juan and {Gaztanaga}, Enrique and {Pietro Gentile Fusillo}, Nicola and {Gerard}, Terry and {Gershkovich}, Irena and {Giannantonio}, Tommaso and {Gillet}, Denis and {Gonzalez-de-Rivera}, Guillermo and {Gonzalez-Perez}, Violeta and {Gott}, Shelby and {Graur}, Or and {Gutierrez}, Gaston and {Guy}, Julien and {Habib}, Salman and {Heetderks}, Henry and {Heetderks}, Ian and {Heitmann}, Katrin and {Hellwing}, Wojciech A. and {Herrera}, David A. and {Ho}, Shirley and {Holland}, Stephen and {Honscheid}, Klaus and {Huff}, Eric and {Hutchinson}, Timothy A. and {Huterer}, Dragan and {Hwang}, Ho Seong and {Illa Laguna}, Joseph Maria and {Ishikawa}, Yuzo and {Jacobs}, Dianna and {Jeffrey}, Niall and {Jelinsky}, Patrick and {Jennings}, Elise and {Jiang}, Linhua and {Jimenez}, Jorge and {Johnson}, Jennifer and {Joyce}, Richard and {Jullo}, Eric and {Juneau}, St{\'e}phanie and {Kama}, Sami and {Karcher}, Armin and {Karkar}, Sonia and {Kehoe}, Robert and {Kennamer}, Noble and {Kent}, Stephen and {Kilbinger}, Martin and {Kim}, Alex G. and {Kirkby}, David and {Kisner}, Theodore and {Kitanidis}, Ellie and {Kneib}, Jean-Paul and {Koposov}, Sergey and {Kovacs}, Eve and {Koyama}, Kazuya and {Kremin}, Anthony and {Kron}, Richard and {Kronig}, Luzius and {Kueter-Young}, Andrea and {Lacey}, Cedric G. and {Lafever}, Robin and {Lahav}, Ofer and {Lambert}, Andrew and {Lampton}, Michael and {Landriau}, Martin and {Lang}, Dustin and {Lauer}, Tod R. and {Le Goff}, Jean-Marc and {Le Guillou}, Laurent and {Le Van Suu}, Auguste and {Lee}, Jae Hyeon and {Lee}, Su-Jeong and {Leitner}, Daniela and {Lesser}, Michael and {Levi}, Michael E. and {L'Huillier}, Benjamin and {Li}, Baojiu and {Liang}, Ming and {Lin}, Huan and {Linder}, Eric and {Loebman}, Sarah R. and {Luki{\'c}}, Zarija and {Ma}, Jun and {MacCrann}, Niall and {Magneville}, Christophe and {Makarem}, Laleh and {Manera}, Marc and {Manser}, Christopher J. and {Marshall}, Robert and {Martini}, Paul and {Massey}, Richard and {Matheson}, Thomas and {McCauley}, Jeremy and {McDonald}, Patrick and {McGreer}, Ian D. and {Meisner}, Aaron and {Metcalfe}, Nigel and {Miller}, Timothy N. and {Miquel}, Ramon and {Moustakas}, John and {Myers}, Adam and {Naik}, Milind and {Newman}, Jeffrey A. and {Nichol}, Robert C. and {Nicola}, Andrina and {Nicolati da Costa}, Luiz and {Nie}, Jundan and {Niz}, Gustavo and {Norberg}, Peder and {Nord}, Brian and {Norman}, Dara and {Nugent}, Peter and {O'Brien}, Thomas and {Oh}, Minji and {Olsen}, Knut A.~G.},
        title = "{The DESI Experiment Part I: Science,Targeting, and Survey Design}",
      journal = {arXiv e-prints},
         year = 2016,
        month = oct,
          eid = {arXiv:1611.00036},
        pages = {arXiv:1611.00036},
          doi = {10.48550/arXiv.1611.00036},
archivePrefix = {arXiv},
       eprint = {1611.00036},
 primaryClass = {astro-ph.IM},
       adsurl = {https://ui.adsabs.harvard.edu/abs/2016arXiv161100036D}
}

@ARTICLE{DESIDR1,
       author = {{DESI Collaboration} and {Abdul-Karim}, M. and {Adame}, A. G. and {Aguado}, D. and {Aguilar}, J. and {Ahlen}, S. and {Alam}, S. and {Aldering}, G. and {Alexander}, D. M. and {Alfarsy}, R. and {Allen}, L. and {Allende Prieto}, C. and {Alves}, O. and {Anand}, A. and {Andrade}, U. and {Armengaud}, E. and {Avila}, S. and {Aviles}, A. and {Awan}, H. and {Bailey}, S. and {Baleato Lizancos}, A. and {Ballester}, O. and {Bault}, A. and {Bautista}, J. and {BenZvi}, S. and {Beraldo e Silva}, L. and {Bermejo-Climent}, J. R. and {Beutler}, F. and {Bianchi}, D. and {Blake}, C. and {Blum}, R. and {Bolton}, A. S. and {Bonici}, M. and {Brieden}, S. and {Brodzeller}, A. and {Brooks}, D. and {Buckley-Geer}, E. and {Burtin}, E. and {Canning}, R. and {Carnero Rosell}, A. and {Carr}, A. and {Carrilho}, P. and {Casas}, L. and {Castander}, F. J. and {Cereskaite}, R. and {Cervantes-Cota}, J. L. and {Chaussidon}, E. and {Chaves-Montero}, J. and {Chen}, S. and {Chen}, X. and {Claybaugh}, T. and {Cole}, S. and {Cooper}, A. P. and {Cousinou}, M.-C. and {Cuceu}, A. and {Davis}, T. M. and {Dawson}, K. S. and {de Belsunce}, R. and {de la Cruz}, R. and {de la Macorra}, A. and {de Mattia}, A. and {Deiosso}, N. and {Della Costa}, J. and {Demina}, R. and {Demirbozan}, U. and {DeRose}, J. and {Dey}, A. and {Dey}, B. and {Ding}, J. and {Ding}, Z. and {Doel}, P. and {Douglass}, K. and {Dowicz}, M. and {Ebina}, H. and {Edelstein}, J. and {Eisenstein}, D. J. and {Elbers}, W. and {Emas}, N. and {Escoffier}, S. and {Fagrelius}, P. and {Fan}, X. and {Fanning}, K. and {Fawcett}, V. A. and {Fern{\'a}ndez-Garc{\'i}a}, E. and {Ferraro}, S. and {Findlay}, N. and {Font-Ribera}, A. and {Forero-Romero}, J. E. and {Forero-S{\'a}nchez}, D. and {Frenk}, C. S. and {G{\"a}nsicke}, B. T. and {Galbany}, L. and {Garc{\'i}a-Bellido}, J. and {Garcia-Quintero}, C. and {Garrison}, L. H. and {Gazta{\~n}aga}, E. and {Gil-Mar{\'i}n}, H. and {Gnedin}, O. Y. and {Gontcho, S. Gontcho A} and {Gonzalez-Morales}, A. X. and {Gonzalez-Perez}, V. and {Gordon}, C. and {Graur}, O. and {Green}, D. and {Gruen}, D. and {Gsponer}, R. and {Guandalin}, C. and {Gutierrez}, G. and {Guy}, J. and {Hahn}, C. and {Han}, J. J. and {Han}, J. and {He}, S. and {Herrera-Alcantar}, H. K. and {Honscheid}, K. and {Hou}, J. and {Howlett}, C. and {Huterer}, D. and {Ir{\v s}i{\v c}}, V. and {Ishak}, M. and {Jacques}, A. and {Jimenez}, J. and {Jing}, Y. P. and {Joachimi}, B. and {Joudaki}, S. and {Joyce}, R. and {Jullo}, E. and {Juneau}, S. and {Kara{\c c}ayl{\i}}, N. G. and {Karim}, T. and {Kehoe}, R. and {Kent}, S. and {Khederlarian}, A. and {Kirkby}, D. and {Kisner}, T. and {Kitaura}, F.-S. and {Kizhuprakkat}, N. and {Kong}, H. and {Koposov}, S. E. and {Kremin}, A. and {Krolewski}, A. and {Lahav}, O. and {Lai}, Y. and {Lamman}, C. and {Lan}, T.-W. and {Landriau}, M. and {Lang}, D. and {Lange}, J. U. and {Lasker}, J. and {Le Goff}, J. M. and {Le Guillou}, L. and {Leauthaud}, A. and {Levi}, M. E. and {Li}, S. and {Li}, T. S. and {Lodha}, K. and {Lokken}, M. and {Luo}, Y. and {Magneville}, C. and {Manera}, M. and {Manser}, C. J. and {Margala}, D. and {Martini}, P. and {Maus}, M. and {McCullough}, J. and {McDonald}, P. and {Medina}, G. E. and {Medina-Varela}, L. and {Meisner}, A. and {Mena-Fern{\'a}ndez}, J. and {Menegas}, A. and {Mezcua}, M. and {Miquel}, R. and {Montero-Camacho}, P. and {Moon}, J. and {Moustakas}, J. and {Mu{\~n}oz-Guti{\'e}rrez}, A. and {Mu{\~n}oz-Santos}, D. and {Myers}, A. D. and {Myles}, J. and {Nadathur}, S. and {Najita}, J. and {Napolitano}, L. and {Newman}, J. A. and {Nikakhtar}, F. and {Nikutta}, R. and {Niz}, G. and {Noriega}, H. E. and {Padmanabhan}, N. and {Paillas}, E. and {Palanque-Delabrouille}, N. and {Palmese}, A. and {Pan}, J. and {Pan}, Z. and {Parkinson}, D. and {Peacock}, J. and {Percival}, W. J. and {P{\'e}rez-Fern{\'a}ndez}, A. and {P{\'e}rez-R{\`a}fols}, I. and {Peterson}, P.},
        title = "{Data Release 1 of the Dark Energy Spectroscopic Instrument}",
      journal = {arXiv e-prints},
         year = 2025,
        month = mar,
          eid = {arXiv:2503.14745},
        pages = {arXiv:2503.14745},
          doi = {10.48550/arXiv.2503.14745},
archivePrefix = {arXiv},
       eprint = {2503.14745},
 primaryClass = {astro-ph.CO},
       adsurl = {https://ui.adsabs.harvard.edu/abs/2025arXiv250314745D}
}

@ARTICLE{dong08,
       author = {{Dong}, Xiaobo and {Wang}, Tinggui and {Wang}, Jianguo and {Yuan}, Weimin and {Zhou}, Hongyan and {Dai}, Haifeng and {Zhang}, Kai},
        title = "{Broad-line Balmer decrements in blue active galactic nuclei}",
      journal = {\mnras},
         year = 2008,
        month = jan,
       volume = {383},
       number = {2},
        pages = {581-592},
          doi = {10.1111/j.1365-2966.2007.12560.x},
archivePrefix = {arXiv},
       eprint = {0710.1458},
 primaryClass = {astro-ph},
       adsurl = {https://ui.adsabs.harvard.edu/abs/2008MNRAS.383..581D}
}

@ARTICLE{dong11,
       author = {{Dong}, Xiao-Bo and {Wang}, Jian-Guo and {Ho}, Luis C. and {Wang}, Ting-Gui and {Fan}, Xiaohui and {Wang}, Huiyuan and {Zhou}, Hongyan and {Yuan}, Weimin},
        title = "{What Controls the Fe II Strength in Active Galactic Nuclei?}",
      journal = {\apj},
         year = 2011,
        month = aug,
       volume = {736},
       number = {2},
          eid = {86},
        pages = {86},
          doi = {10.1088/0004-637X/736/2/86},
archivePrefix = {arXiv},
       eprint = {0903.5020},
 primaryClass = {astro-ph.GA},
       adsurl = {https://ui.adsabs.harvard.edu/abs/2011ApJ...736...86D}
}

@ARTICLE{dong12,
       author = {{Dong}, Xiao-Bo and {Ho}, Luis C. and {Yuan}, Weimin and {Wang}, Ting-Gui and {Fan}, Xiaohui and {Zhou}, Hongyan and {Jiang}, Ning},
        title = "{A Uniformly Selected Sample of Low-mass Black Holes in Seyfert 1 Galaxies}",
      journal = {\apj},
         year = 2012,
        month = aug,
       volume = {755},
       number = {2},
          eid = {167},
        pages = {167},
          doi = {10.1088/0004-637X/755/2/167},
archivePrefix = {arXiv},
       eprint = {1206.3843},
 primaryClass = {astro-ph.GA},
       adsurl = {https://ui.adsabs.harvard.edu/abs/2012ApJ...755..167D}
}

@ARTICLE{Fanidakis2012,
       author = {{Fanidakis}, N. and {Baugh}, C.~M. and {Benson}, A.~J. and {Bower}, R.~G. and {Cole}, S. and {Done}, C. and {Frenk}, C.~S. and {Hickox}, R.~C. and {Lacey}, C. and {Lagos}, C. del P.},
        title = "{The evolution of active galactic nuclei across cosmic time: what is downsizing?}",
      journal = {\mnras},
         year = 2012,
        month = feb,
       volume = {419},
       number = {4},
        pages = {2797-2820},
          doi = {10.1111/j.1365-2966.2011.19931.x},
archivePrefix = {arXiv},
       eprint = {1011.5222},
 primaryClass = {astro-ph.CO},
       adsurl = {https://ui.adsabs.harvard.edu/abs/2012MNRAS.419.2797F}
}

@ARTICLE{NGC4395_Filippenko93,
       author = {{Filippenko}, Alexei V. and {Ho}, Luis C. and {Sargent}, Wallace L.~W.},
        title = "{HST Observations of NGC 4395, the Least Luminous Seyfert 1 Nucleus: Evidence against the Starburst Hypothesis for Broad-lined Active Galactic Nuclei}",
      journal = {\apjl},
         year = 1993,
        month = jun,
       volume = {410},
        pages = {L75},
          doi = {10.1086/186883},
       adsurl = {https://ui.adsabs.harvard.edu/abs/1993ApJ...410L..75F}
}

@ARTICLE{NGC4395_Filippenko2003,
       author = {{Filippenko}, Alexei V. and {Ho}, Luis C.},
        title = "{A Low-Mass Central Black Hole in the Bulgeless Seyfert 1 Galaxy NGC 4395}",
      journal = {\apjl},
         year = 2003,
        month = may,
       volume = {588},
       number = {1},
        pages = {L13-L16},
          doi = {10.1086/375361},
archivePrefix = {arXiv},
       eprint = {astro-ph/0303429},
 primaryClass = {astro-ph},
       adsurl = {https://ui.adsabs.harvard.edu/abs/2003ApJ...588L..13F}
}

@ARTICLE{Fitzpatrick1999,
       author = {{Fitzpatrick}, Edward L.},
        title = "{Correcting for the Effects of Interstellar Extinction}",
      journal = {\pasp},
         year = 1999,
        month = jan,
       volume = {111},
       number = {755},
        pages = {63-75},
          doi = {10.1086/316293},
archivePrefix = {arXiv},
       eprint = {astro-ph/9809387},
 primaryClass = {astro-ph},
       adsurl = {https://ui.adsabs.harvard.edu/abs/1999PASP..111...63F}
}

@ARTICLE{Fontanot2020,
       author = {{Fontanot}, Fabio and {De Lucia}, Gabriella and {Hirschmann}, Michaela and {Xie}, Lizhi and {Monaco}, Pierluigi and {Menci}, Nicola and {Fiore}, Fabrizio and {Feruglio}, Chiara and {Cristiani}, Stefano and {Shankar}, Francesco},
        title = "{The rise of active galactic nuclei in the galaxy evolution and assembly semi-analytic model}",
      journal = {\mnras},
         year = 2020,
        month = aug,
       volume = {496},
       number = {3},
        pages = {3943-3960},
          doi = {10.1093/mnras/staa1716},
archivePrefix = {arXiv},
       eprint = {2002.10576},
 primaryClass = {astro-ph.CO},
       adsurl = {https://ui.adsabs.harvard.edu/abs/2020MNRAS.496.3943F}
}

@ARTICLE{gh04,
       author = {{Greene}, Jenny E. and {Ho}, Luis C.},
        title = "{Active Galactic Nuclei with Candidate Intermediate-Mass Black Holes}",
      journal = {\apj},
         year = 2004,
        month = aug,
       volume = {610},
       number = {2},
        pages = {722-736},
          doi = {10.1086/421719},
archivePrefix = {arXiv},
       eprint = {astro-ph/0404110},
 primaryClass = {astro-ph},
       adsurl = {https://ui.adsabs.harvard.edu/abs/2004ApJ...610..722G}
}

@ARTICLE{gh05,
       author = {{Greene}, Jenny E. and {Ho}, Luis C.},
        title = "{Estimating Black Hole Masses in Active Galaxies Using the H{\ensuremath{\alpha}} Emission Line}",
      journal = {\apj},
         year = 2005,
        month = sep,
       volume = {630},
       number = {1},
        pages = {122-129},
          doi = {10.1086/431897},
archivePrefix = {arXiv},
       eprint = {astro-ph/0508335},
 primaryClass = {astro-ph},
       adsurl = {https://ui.adsabs.harvard.edu/abs/2005ApJ...630..122G}
}

@ARTICLE{gh07,
       author = {{Greene}, Jenny E. and {Ho}, Luis C.},
        title = "{A New Sample of Low-Mass Black Holes in Active Galaxies}",
      journal = {\apj},
         year = 2007,
        month = nov,
       volume = {670},
       number = {1},
        pages = {92-104},
          doi = {10.1086/522082},
archivePrefix = {arXiv},
       eprint = {0707.2617},
 primaryClass = {astro-ph},
       adsurl = {https://ui.adsabs.harvard.edu/abs/2007ApJ...670...92G}
}

@article{gh07MF,
       author = {{Greene}, Jenny E. and {Ho}, Luis C.},
        title = "{The Mass Function of Active Black Holes in the Local Universe}",
      journal = {\apj},
         year = 2007,
        month = sep,
       volume = {667},
       number = {1},
        pages = {131-148},
          doi = {10.1086/520497},
archivePrefix = {arXiv},
       eprint = {0705.0020},
 primaryClass = {astro-ph},
       adsurl = {https://ui.adsabs.harvard.edu/abs/2007ApJ...667..131G}
}

@ARTICLE{greene_araa,
       author = {{Greene}, Jenny E. and {Strader}, Jay and {Ho}, Luis C.},
        title = "{Intermediate-Mass Black Holes}",
      journal = {\araa},
         year = 2020,
        month = aug,
       volume = {58},
        pages = {257-312},
          doi = {10.1146/annurev-astro-032620-021835},
archivePrefix = {arXiv},
       eprint = {1911.09678},
 primaryClass = {astro-ph.GA},
       adsurl = {https://ui.adsabs.harvard.edu/abs/2020ARA&A..58..257G}
}

@ARTICLE{Hasinger2005,
       author = {{Hasinger}, G. and {Miyaji}, T. and {Schmidt}, M.},
        title = "{Luminosity-dependent evolution of soft X-ray selected AGN. New Chandra and XMM-Newton surveys}",
      journal = {\aap},
         year = 2005,
        month = oct,
       volume = {441},
       number = {2},
        pages = {417-434},
          doi = {10.1051/0004-6361:20042134},
archivePrefix = {arXiv},
       eprint = {astro-ph/0506118},
 primaryClass = {astro-ph},
       adsurl = {https://ui.adsabs.harvard.edu/abs/2005A&A...441..417H}
}

@ARTICLE{heckman1980,
       author = {{Heckman}, T.~M.},
        title = "{An Optical and Radio Survey of the Nuclei of Bright Galaxies - Activity in the Normal Galactic Nuclei}",
      journal = {\aap},
         year = 1980,
        month = jul,
       volume = {87},
        pages = {152},
       adsurl = {https://ui.adsabs.harvard.edu/abs/1980A&A....87..152H}
}

@article{heckman04,
       author = {{Heckman}, Timothy M. and {Kauffmann}, Guinevere and {Brinchmann}, Jarle and {Charlot}, St{\'e}phane and {Tremonti}, Christy and {White}, Simon D.~M.},
        title = "{Present-Day Growth of Black Holes and Bulges: The Sloan Digital Sky Survey Perspective}",
      journal = {Astrophys. J.},
         year = 2004,
        month = sep,
       volume = {613},
       number = {1},
        pages = {109-118},
          doi = {10.1086/422872},
archivePrefix = {arXiv},
       eprint = {astro-ph/0406218},
 primaryClass = {astro-ph},
       adsurl = {https://ui.adsabs.harvard.edu/abs/2004ApJ...613..109H}
}

@ARTICLE{Hirschmann2014,
       author = {{Hirschmann}, Michaela and {Dolag}, Klaus and {Saro}, Alexandro and {Bachmann}, Lisa and {Borgani}, Stefano and {Burkert}, Andreas},
        title = "{Cosmological simulations of black hole growth: AGN luminosities and downsizing}",
      journal = {\mnras},
         year = 2014,
        month = aug,
       volume = {442},
       number = {3},
        pages = {2304-2324},
          doi = {10.1093/mnras/stu1023},
archivePrefix = {arXiv},
       eprint = {1308.0333},
 primaryClass = {astro-ph.CO},
       adsurl = {https://ui.adsabs.harvard.edu/abs/2014MNRAS.442.2304H}
}

@ARTICLE{Ho1997a,
       author = {{Ho}, Luis C. and {Filippenko}, Alexei V. and {Sargent}, Wallace L.~W.},
        title = "{A Search for ``Dwarf'' Seyfert Nuclei. III. Spectroscopic Parameters and Properties of the Host Galaxies}",
      journal = {\apjs},
         year = 1997,
        month = oct,
       volume = {112},
       number = {2},
        pages = {315-390},
          doi = {10.1086/313041},
archivePrefix = {arXiv},
       eprint = {astro-ph/9704107},
 primaryClass = {astro-ph},
       adsurl = {https://ui.adsabs.harvard.edu/abs/1997ApJS..112..315H}
}

@ARTICLE{Ho1997b,
       author = {{Ho}, Luis C. and {Filippenko}, Alexei V. and {Sargent}, Wallace L.~W. and {Peng}, Chien Y.},
        title = "{A Search for ``Dwarf'' Seyfert Nuclei. IV. Nuclei with Broad H{\ensuremath{\alpha}} Emission}",
      journal = {\apjs},
         year = 1997,
        month = oct,
       volume = {112},
       number = {2},
        pages = {391-414},
          doi = {10.1086/313042},
archivePrefix = {arXiv},
       eprint = {astro-ph/9704099},
 primaryClass = {astro-ph},
       adsurl = {https://ui.adsabs.harvard.edu/abs/1997ApJS..112..391H}
}

@ARTICLE{Ho1997c,
       author = {{Ho}, Luis C. and {Filippenko}, Alexei V. and {Sargent}, Wallace L.~W.},
        title = "{A Search for ``Dwarf'' Seyfert Nuclei. V. Demographics of Nuclear Activity in Nearby Galaxies}",
      journal = {\apj},
         year = 1997,
        month = oct,
       volume = {487},
       number = {2},
        pages = {568-578},
          doi = {10.1086/304638},
archivePrefix = {arXiv},
       eprint = {astro-ph/9704108},
 primaryClass = {astro-ph},
       adsurl = {https://ui.adsabs.harvard.edu/abs/1997ApJ...487..568H}
}

@ARTICLE{Ho2008,
       author = {{Ho}, L.~C.},
        title = "{Nuclear activity in nearby galaxies.}",
      journal = {\araa},
         year = 2008,
        month = sep,
       volume = {46},
        pages = {475-539},
          doi = {10.1146/annurev.astro.45.051806.110546},
archivePrefix = {arXiv},
       eprint = {0803.2268},
 primaryClass = {astro-ph},
       adsurl = {https://ui.adsabs.harvard.edu/abs/2008ARA&A..46..475H}
}

@ARTICLE{Ho2009,
       author = {{Ho}, Luis C.},
        title = "{Radiatively Inefficient Accretion in Nearby Galaxies}",
      journal = {\apj},
         year = 2009,
        month = jul,
       volume = {699},
       number = {1},
        pages = {626-637},
          doi = {10.1088/0004-637X/699/1/626},
archivePrefix = {arXiv},
       eprint = {0906.4104},
 primaryClass = {astro-ph.GA},
       adsurl = {https://ui.adsabs.harvard.edu/abs/2009ApJ...699..626H}
}

@ARTICLE{Ho&Kim2014,
       author = {{Ho}, Luis C. and {Kim}, Minjin},
        title = "{The Black Hole Mass Scale of Classical and Pseudo Bulges in Active Galaxies}",
      journal = {\apj},
         year = 2014,
        month = jul,
       volume = {789},
       number = {1},
          eid = {17},
        pages = {17},
          doi = {10.1088/0004-637X/789/1/17},
archivePrefix = {arXiv},
       eprint = {1406.6137},
 primaryClass = {astro-ph.GA},
       adsurl = {https://ui.adsabs.harvard.edu/abs/2014ApJ...789...17H}
}

@ARTICLE{Ho_Kim2015,
       author = {{Ho}, Luis C. and {Kim}, Minjin},
        title = "{A Revised Calibration of the Virial Mass Estimator for Black Holes in Active Galaxies Based on Single-epoch H{\ensuremath{\beta}} Spectra}",
      journal = {\apj},
         year = 2015,
        month = aug,
       volume = {809},
       number = {2},
          eid = {123},
        pages = {123},
          doi = {10.1088/0004-637X/809/2/123},
archivePrefix = {arXiv},
       eprint = {1507.04821},
 primaryClass = {astro-ph.GA},
       adsurl = {https://ui.adsabs.harvard.edu/abs/2015ApJ...809..123H}
}

@ARTICLE{Inayoshi2020,
       author = {{Inayoshi}, Kohei and {Visbal}, Eli and {Haiman}, Zolt{\'a}n},
        title = "{The Assembly of the First Massive Black Holes}",
      journal = {\araa},
         year = 2020,
        month = aug,
       volume = {58},
        pages = {27-97},
          doi = {10.1146/annurev-astro-120419-014455},
archivePrefix = {arXiv},
       eprint = {1911.05791},
 primaryClass = {astro-ph.GA},
       adsurl = {https://ui.adsabs.harvard.edu/abs/2020ARA&A..58...27I}
}

@ARTICLE{NGC4395_Kraemer99,
       author = {{Kraemer}, Steven B. and {Ho}, Luis C. and {Crenshaw}, D. Michael and {Shields}, Joseph C. and {Filippenko}, Alexei V.},
        title = "{Physical Conditions in the Emission-Line Gas in the Extremely Low Luminosity Seyfert Nucleus of NGC 4395}",
      journal = {\apj},
         year = 1999,
        month = aug,
       volume = {520},
       number = {2},
        pages = {564-573},
          doi = {10.1086/307486},
archivePrefix = {arXiv},
       eprint = {astro-ph/9902163},
 primaryClass = {astro-ph},
       adsurl = {https://ui.adsabs.harvard.edu/abs/1999ApJ...520..564K}
}

@ARTICLE{kewley01,
       author = {{Kewley}, L.~J. and {Dopita}, M.~A. and {Sutherland}, R.~S. and {Heisler}, C.~A. and {Trevena}, J.},
        title = "{Theoretical Modeling of Starburst Galaxies}",
      journal = {\apj},
         year = 2001,
        month = jul,
       volume = {556},
       number = {1},
        pages = {121-140},
          doi = {10.1086/321545},
archivePrefix = {arXiv},
       eprint = {astro-ph/0106324},
 primaryClass = {astro-ph},
       adsurl = {https://ui.adsabs.harvard.edu/abs/2001ApJ...556..121K}
}

@ARTICLE{kewley06,
       author = {{Kewley}, Lisa J. and {Groves}, Brent and {Kauffmann}, Guinevere and {Heckman}, Tim},
        title = "{The host galaxies and classification of active galactic nuclei}",
      journal = {\mnras},
         year = 2006,
        month = nov,
       volume = {372},
       number = {3},
        pages = {961-976},
          doi = {10.1111/j.1365-2966.2006.10859.x},
archivePrefix = {arXiv},
       eprint = {astro-ph/0605681},
 primaryClass = {astro-ph},
       adsurl = {https://ui.adsabs.harvard.edu/abs/2006MNRAS.372..961K}
}

@ARTICLE{Kauffmann03a,
       author = {{Kauffmann}, Guinevere and {Heckman}, Timothy M. and {White}, Simon D.~M. and {Charlot}, St{\'e}phane and {Tremonti}, Christy and {Brinchmann}, Jarle and {Bruzual}, Gustavo and {Peng}, Eric W. and {Seibert}, Mark and {Bernardi}, Mariangela and {Blanton}, Michael and {Brinkmann}, Jon and {Castander}, Francisco and {Cs{\'a}bai}, Istvan and {Fukugita}, Masataka and {Ivezic}, Zeljko and {Munn}, Jeffrey A. and {Nichol}, Robert C. and {Padmanabhan}, Nikhil and {Thakar}, Aniruddha R. and {Weinberg}, David H. and {York}, Donald},
        title = "{Stellar masses and star formation histories for {}10$^{5}$ galaxies from the Sloan Digital Sky Survey}",
      journal = {\mnras},
         year = 2003,
        month = may,
       volume = {341},
       number = {1},
        pages = {33-53},
          doi = {10.1046/j.1365-8711.2003.06291.x},
archivePrefix = {arXiv},
       eprint = {astro-ph/0204055},
 primaryClass = {astro-ph},
       adsurl = {https://ui.adsabs.harvard.edu/abs/2003MNRAS.341...33K}
}

@ARTICLE{kauffmann03b,
       author = {{Kauffmann}, Guinevere and {Heckman}, Timothy M. and {Tremonti}, Christy and {Brinchmann}, Jarle and {Charlot}, St{\'e}phane and {White}, Simon D.~M. and {Ridgway}, Susan E. and {Brinkmann}, Jon and {Fukugita}, Masataka and {Hall}, Patrick B. and {Ivezi{\'c}}, {\v{Z}}eljko and {Richards}, Gordon T. and {Schneider}, Donald P.},
        title = "{The host galaxies of active galactic nuclei}",
      journal = {\mnras},
         year = 2003,
        month = dec,
       volume = {346},
       number = {4},
        pages = {1055-1077},
          doi = {10.1111/j.1365-2966.2003.07154.x},
archivePrefix = {arXiv},
       eprint = {astro-ph/0304239},
 primaryClass = {astro-ph},
       adsurl = {https://ui.adsabs.harvard.edu/abs/2003MNRAS.346.1055K}
}

@ARTICLE{Kormendy&Ho_araa,
       author = {{Kormendy}, John and {Ho}, Luis C.},
        title = "{Coevolution (Or Not) of Supermassive Black Holes and Host Galaxies}",
      journal = {\araa},
         year = 2013,
        month = aug,
       volume = {51},
       number = {1},
        pages = {511-653},
          doi = {10.1146/annurev-astro-082708-101811},
archivePrefix = {arXiv},
       eprint = {1304.7762},
 primaryClass = {astro-ph.CO},
       adsurl = {https://ui.adsabs.harvard.edu/abs/2013ARA&A..51..511K}
}

@ARTICLE{Laor1997,
       author = {{Laor}, Ari and {Jannuzi}, Buell T. and {Green}, Richard F. and {Boroson}, Todd A.},
        title = "{The Ultraviolet Properties of the Narrow-Line Quasar I Zw 1}",
      journal = {\apj},
         year = 1997,
        month = nov,
       volume = {489},
       number = {2},
        pages = {656-671},
          doi = {10.1086/304816},
archivePrefix = {arXiv},
       eprint = {astro-ph/9706264},
 primaryClass = {astro-ph},
       adsurl = {https://ui.adsabs.harvard.edu/abs/1997ApJ...489..656L}
}

@ARTICLE{Le&Woo2020,
       author = {{Le}, Huynh Anh N. and {Woo}, Jong-Hak and {Xue}, Yongquan},
        title = "{Calibrating Mg II-based Black Hole Mass Estimators Using Low-to-high-luminosity Active Galactic Nuclei}",
      journal = {\apj},
         year = 2020,
        month = sep,
       volume = {901},
       number = {1},
          eid = {35},
        pages = {35},
          doi = {10.3847/1538-4357/abada0},
archivePrefix = {arXiv},
       eprint = {2008.02990},
 primaryClass = {astro-ph.GA},
       adsurl = {https://ui.adsabs.harvard.edu/abs/2020ApJ...901...35L}
}

@ARTICLE{j0838,
       author = {{Liu}, Wen-Juan and {Qian}, Lei and {Dong}, Xiao-Bo and {Jiang}, Ning and {Lira}, Paulina and {Cai}, Zheng and {Wang}, Feige and {Yang}, Jinyi and {Xiao}, Ting and {Kim}, Minjin},
        title = "{A Ringed Dwarf LINER 1 Galaxy Hosting an Intermediate-mass Black Hole with Large-scale Rotation-like H{\ensuremath{\alpha}} Emission}",
      journal = {\apj},
         year = 2017,
        month = mar,
       volume = {837},
       number = {2},
          eid = {109},
        pages = {109},
          doi = {10.3847/1538-4357/aa5eb6},
archivePrefix = {arXiv},
       eprint = {1702.01033},
 primaryClass = {astro-ph.GA},
       adsurl = {https://ui.adsabs.harvard.edu/abs/2017ApJ...837..109L}
}

@ARTICLE{liu18,
       author = {{Liu}, He-Yang and {Yuan}, Weimin and {Dong}, Xiao-Bo and {Zhou}, Hongyan and {Liu}, Wen-Juan},
        title = "{A Uniformly Selected Sample of Low-mass Black Holes in Seyfert 1 Galaxies. II. The SDSS DR7 Sample}",
      journal = {\apjs},
         year = 2018,
        month = apr,
       volume = {235},
       number = {2},
          eid = {40},
        pages = {40},
          doi = {10.3847/1538-4365/aab88e},
archivePrefix = {arXiv},
       eprint = {1803.04330},
 primaryClass = {astro-ph.GA},
       adsurl = {https://ui.adsabs.harvard.edu/abs/2018ApJS..235...40L}
}

@ARTICLE{liu19,
       author = {{Liu}, He-Yang and {Liu}, Wen-Juan and {Dong}, Xiao-Bo and {Zhou}, Hongyan and {Wang}, Tinggui and {Lu}, Honglin and {Yuan}, Weimin},
        title = "{A Comprehensive and Uniform Sample of Broad-line Active Galactic Nuclei from the SDSS DR7}",
      journal = {\apjs},
         year = 2019,
        month = aug,
       volume = {243},
       number = {2},
          eid = {21},
        pages = {21},
          doi = {10.3847/1538-4365/ab298b},
archivePrefix = {arXiv},
       eprint = {1906.05597},
 primaryClass = {astro-ph.GA},
       adsurl = {https://ui.adsabs.harvard.edu/abs/2019ApJS..243...21L}
}

@ARTICLE{liu2021,
       author = {{Liu}, Wen-Juan and {Lira}, Paulina and {Yao}, Su and {Xu}, Dawei and {Wang}, Jing and {Dong}, Xiao-Bo and {Mart{\'\i}nez-Palomera}, Jorge},
        title = "{Local Active Galactic Nuclei with Large Broad-H{\ensuremath{\alpha}} Variability Reside in Red Galaxies}",
      journal = {\apj},
         year = 2021,
        month = jul,
       volume = {915},
       number = {1},
          eid = {63},
        pages = {63},
          doi = {10.3847/1538-4357/abf82c},
archivePrefix = {arXiv},
       eprint = {2103.11935},
 primaryClass = {astro-ph.GA},
       adsurl = {https://ui.adsabs.harvard.edu/abs/2021ApJ...915...63L}
}

@ARTICLE{lu2006,
       author = {{Lu}, Honglin and {Zhou}, Hongyan and {Wang}, Junxian and {Wang}, Tinggui and {Dong}, Xiaobo and {Zhuang}, Zhenquan and {Li}, Cheng},
        title = "{Ensemble Learning for Independent Component Analysis of Normal Galaxy Spectra}",
      journal = {\aj},
         year = 2006,
        month = feb,
       volume = {131},
       number = {2},
        pages = {790-805},
          doi = {10.1086/498711},
archivePrefix = {arXiv},
       eprint = {astro-ph/0510246},
 primaryClass = {astro-ph},
       adsurl = {https://ui.adsabs.harvard.edu/abs/2006AJ....131..790L}
}

@ARTICLE{GN-z11,
       author = {{Maiolino}, Roberto and {Scholtz}, Jan and {Witstok}, Joris and {Carniani}, Stefano and {D'Eugenio}, Francesco and {de Graaff}, Anna and {{\"U}bler}, Hannah and {Tacchella}, Sandro and {Curtis-Lake}, Emma and {Arribas}, Santiago and {Bunker}, Andrew and {Charlot}, St{\'e}phane and {Chevallard}, Jacopo and {Curti}, Mirko and {Looser}, Tobias J. and {Maseda}, Michael V. and {Rawle}, Timothy D. and {Rodr{\'\i}guez del Pino}, Bruno and {Willott}, Chris J. and {Egami}, Eiichi and {Eisenstein}, Daniel J. and {Hainline}, Kevin N. and {Robertson}, Brant and {Williams}, Christina C. and {Willmer}, Christopher N.~A. and {Baker}, William M. and {Boyett}, Kristan and {DeCoursey}, Christa and {Fabian}, Andrew C. and {Helton}, Jakob M. and {Ji}, Zhiyuan and {Jones}, Gareth C. and {Kumari}, Nimisha and {Laporte}, Nicolas and {Nelson}, Erica J. and {Perna}, Michele and {Sandles}, Lester and {Shivaei}, Irene and {Sun}, Fengwu},
        title = "{A small and vigorous black hole in the early Universe}",
      journal = {\nat},
         year = 2024,
        month = mar,
       volume = {627},
       number = {8002},
        pages = {59-63},
          doi = {10.1038/s41586-024-07052-5},
archivePrefix = {arXiv},
       eprint = {2305.12492},
 primaryClass = {astro-ph.GA},
       adsurl = {https://ui.adsabs.harvard.edu/abs/2024Natur.627...59M}
}

@INPROCEEDINGS{mpfit,
       author = {{Markwardt}, C.~B.},
        title = "{Non-linear Least-squares Fitting in IDL with MPFIT}",
    booktitle = {Astronomical Data Analysis Software and Systems XVIII},
         year = 2009,
       editor = {{Bohlender}, D.~A. and {Durand}, D. and {Dowler}, P.},
       series = {Astronomical Society of the Pacific Conference Series},
       volume = {411},
        month = sep,
        pages = {251},
          doi = {10.48550/arXiv.0902.2850},
archivePrefix = {arXiv},
       eprint = {0902.2850},
 primaryClass = {astro-ph.IM},
       adsurl = {https://ui.adsabs.harvard.edu/abs/2009ASPC..411..251M}
}

@INPROCEEDINGS{mage,
       author = {{Marshall}, J.~L. and {Burles}, Scott and {Thompson}, Ian B. and {Shectman}, Stephen A. and {Bigelow}, Bruce C. and {Burley}, Gregory and {Birk}, Christoph and {Estrada}, Jorge and {Jones}, Patricio and {Smith}, Matthew and {Kowal}, Vince and {Castillo}, Jerson and {Storts}, Robert and {Ortiz}, Greg},
        title = "{The MagE spectrograph}",
    booktitle = {Ground-based and Airborne Instrumentation for Astronomy II},
         year = 2008,
       editor = {{McLean}, Ian S. and {Casali}, Mark M.},
       series = {Society of Photo-Optical Instrumentation Engineers (SPIE) Conference Series},
       volume = {7014},
        month = jul,
          eid = {701454},
        pages = {701454},
          doi = {10.1117/12.789972},
archivePrefix = {arXiv},
       eprint = {0807.3774},
 primaryClass = {astro-ph},
       adsurl = {https://ui.adsabs.harvard.edu/abs/2008SPIE.7014E..54M}
}

@ARTICLE{merloni2008,
       author = {{Merloni}, Andrea and {Heinz}, Sebastian},
        title = "{A synthesis model for AGN evolution: supermassive black holes growth and feedback modes}",
      journal = {\mnras},
         year = 2008,
        month = aug,
       volume = {388},
       number = {3},
        pages = {1011-1030},
          doi = {10.1111/j.1365-2966.2008.13472.x},
archivePrefix = {arXiv},
       eprint = {0805.2499},
 primaryClass = {astro-ph},
       adsurl = {https://ui.adsabs.harvard.edu/abs/2008MNRAS.388.1011M}
}

@ARTICLE{Pucha2025,
       author = {{Pucha}, Ragadeepika and {Juneau}, S. and {Dey}, Arjun and {Siudek}, M. and {Mezcua}, M. and {Moustakas}, J. and {BenZvi}, S. and {Hainline}, K. and {Hviding}, R. and {Mao}, Yao-Yuan and {Alexander}, D.~M. and {Alfarsy}, R. and {Circosta}, C. and {Guo}, Wei-Jian and {Manwadkar}, V. and {Martini}, P. and {Weaver}, B.~A. and {Aguilar}, J. and {Ahlen}, S. and {Bianchi}, D. and {Brooks}, D. and {Canning}, R. and {Claybaugh}, T. and {Dawson}, K. and {de la Macorra}, A. and {Dey}, Biprateep and {Doel}, P. and {Font-Ribera}, A. and {Forero-Romero}, J.~E. and {Gazta{\~n}aga}, E. and {Gontcho A Gontcho}, S. and {Gutierrez}, G. and {Honscheid}, K. and {Kehoe}, R. and {Koposov}, S.~E. and {Lambert}, A. and {Landriau}, M. and {Le Guillou}, L. and {Meisner}, A. and {Miquel}, R. and {Prada}, F. and {Rossi}, G. and {Sanchez}, E. and {Schlegel}, D. and {Schubnell}, M. and {Seo}, H. and {Sprayberry}, D. and {Tarl{\'e}}, G. and {Zou}, H.},
        title = "{Tripling the Census of Dwarf AGN Candidates Using DESI Early Data}",
      journal = {\apj},
         year = 2025,
        month = mar,
       volume = {982},
       number = {1},
          eid = {10},
        pages = {10},
          doi = {10.3847/1538-4357/adb1dd},
archivePrefix = {arXiv},
       eprint = {2411.00091},
 primaryClass = {astro-ph.GA},
       adsurl = {https://ui.adsabs.harvard.edu/abs/2025ApJ...982...10P}
}

@ARTICLE{Reines2013,
       author = {{Reines}, Amy E. and {Greene}, Jenny E. and {Geha}, Marla},
        title = "{Dwarf Galaxies with Optical Signatures of Active Massive Black Holes}",
      journal = {\apj},
         year = 2013,
        month = oct,
       volume = {775},
       number = {2},
          eid = {116},
        pages = {116},
          doi = {10.1088/0004-637X/775/2/116},
archivePrefix = {arXiv},
       eprint = {1308.0328},
 primaryClass = {astro-ph.CO},
       adsurl = {https://ui.adsabs.harvard.edu/abs/2013ApJ...775..116R}
}

@ARTICLE{runnoe12,
       author = {{Runnoe}, Jessie C. and {Brotherton}, Michael S. and {Shang}, Zhaohui},
        title = "{Updating quasar bolometric luminosity corrections}",
      journal = {\mnras},
         year = 2012,
        month = may,
       volume = {422},
       number = {1},
        pages = {478-493},
          doi = {10.1111/j.1365-2966.2012.20620.x},
archivePrefix = {arXiv},
       eprint = {1201.5155},
 primaryClass = {astro-ph.CO},
       adsurl = {https://ui.adsabs.harvard.edu/abs/2012MNRAS.422..478R}
}

@ARTICLE{Schlegel1998,
       author = {{Schlegel}, David J. and {Finkbeiner}, Douglas P. and {Davis}, Marc},
        title = "{Maps of Dust Infrared Emission for Use in Estimation of Reddening and Cosmic Microwave Background Radiation Foregrounds}",
      journal = {\apj},
         year = 1998,
        month = jun,
       volume = {500},
       number = {2},
        pages = {525-553},
          doi = {10.1086/305772},
archivePrefix = {arXiv},
       eprint = {astro-ph/9710327},
 primaryClass = {astro-ph},
       adsurl = {https://ui.adsabs.harvard.edu/abs/1998ApJ...500..525S}
}

@ARTICLE{shankar2013,
       author = {{Shankar}, Francesco and {Weinberg}, David H. and {Miralda-Escud{\'e}}, Jordi},
        title = "{Accretion-driven evolution of black holes: Eddington ratios, duty cycles and active galaxy fractions}",
      journal = {\mnras},
         year = 2013,
        month = jan,
       volume = {428},
       number = {1},
        pages = {421-446},
          doi = {10.1093/mnras/sts026},
archivePrefix = {arXiv},
       eprint = {1111.3574},
 primaryClass = {astro-ph.CO},
       adsurl = {https://ui.adsabs.harvard.edu/abs/2013MNRAS.428..421S}
}

@ARTICLE{shen&kelly2012,
       author = {{Shen}, Yue and {Kelly}, Brandon C.},
        title = "{The Demographics of Broad-line Quasars in the Mass-Luminosity Plane. I. Testing FWHM-based Virial Black Hole Masses}",
      journal = {\apj},
         year = 2012,
        month = feb,
       volume = {746},
       number = {2},
          eid = {169},
        pages = {169},
          doi = {10.1088/0004-637X/746/2/169},
archivePrefix = {arXiv},
       eprint = {1107.4372},
 primaryClass = {astro-ph.CO},
       adsurl = {https://ui.adsabs.harvard.edu/abs/2012ApJ...746..169S}
}

@ARTICLE{sijacki2015,
       author = {{Sijacki}, Debora and {Vogelsberger}, Mark and {Genel}, Shy and {Springel}, Volker and {Torrey}, Paul and {Snyder}, Gregory F. and {Nelson}, Dylan and {Hernquist}, Lars},
        title = "{The Illustris simulation: the evolving population of black holes across cosmic time}",
      journal = {\mnras},
         year = 2015,
        month = sep,
       volume = {452},
       number = {1},
        pages = {575-596},
          doi = {10.1093/mnras/stv1340},
archivePrefix = {arXiv},
       eprint = {1408.6842},
 primaryClass = {astro-ph.GA},
       adsurl = {https://ui.adsabs.harvard.edu/abs/2015MNRAS.452..575S}
}

@ARTICLE{Sun2025,
       author = {{Sun}, Jingbo and {Guo}, Hengxiao and {Zuo}, Wenwen and {Lira}, Paulina and {Gu}, Minfeng and {Edwards}, Philip G. and {Wang}, Shu and {Stevens}, Jamie and {An}, Tao and {Barua}, Samuzal and {Cai}, Zhen-yi and {Feng}, Haicheng and {Gupta}, Alok C. and {Ho}, Luis C. and {Ili{\'c}}, Dragana and {Kova{\v{c}}evi{\'c}}, Andjelka B. and {Li}, ShaSha and {Mezcua}, Mar and {Popovi{\'c}}, Luka {\v{C}}. and {S{\'a}nchez-S{\'a}ez}, Paula and {Sun}, Mouyuan and {Shen}, Rongfeng and {U}, Vivian and {Vince}, Oliver and {Wang}, Junxian and {Wu}, Xuebing and {Yu}, Zhefu and {Zheng}, Zhenya},
        title = "{The Intermediate-mass Black Hole Reverberation Mapping Project: First Detection of Mid-infrared Lags in Prototypical Intermediate-mass Black Holes in NGC 4395 and POX 52}",
      journal = {\apjl},
         year = 2025,
        month = aug,
       volume = {989},
       number = {2},
          eid = {L26},
        pages = {L26},
          doi = {10.3847/2041-8213/adf3a3},
archivePrefix = {arXiv},
       eprint = {2504.21711},
 primaryClass = {astro-ph.GA},
       adsurl = {https://ui.adsabs.harvard.edu/abs/2025ApJ...989L..26S}
}

@ARTICLE{uvFeII_T06,
       author = {{Tsuzuki}, Yumihiko and {Kawara}, Kimiaki and {Yoshii}, Yuzuru and {Oyabu}, Shinki and {Tanab{\'e}}, Toshihiko and {Matsuoka}, Yoshiki},
        title = "{Fe II Emission in 14 Low-Redshift Quasars. I. Observations}",
      journal = {\apj},
         year = 2006,
        month = oct,
       volume = {650},
       number = {1},
        pages = {57-79},
          doi = {10.1086/506376},
archivePrefix = {arXiv},
       eprint = {astro-ph/0606040},
 primaryClass = {astro-ph},
       adsurl = {https://ui.adsabs.harvard.edu/abs/2006ApJ...650...57T}
}

@ARTICLE{Ueda2003,
       author = {{Ueda}, Yoshihiro and {Akiyama}, Masayuki and {Ohta}, Kouji and {Miyaji}, Takamitsu},
        title = "{Cosmological Evolution of the Hard X-Ray Active Galactic Nucleus Luminosity Function and the Origin of the Hard X-Ray Background}",
      journal = {\apj},
         year = 2003,
        month = dec,
       volume = {598},
       number = {2},
        pages = {886-908},
          doi = {10.1086/378940},
archivePrefix = {arXiv},
       eprint = {astro-ph/0308140},
 primaryClass = {astro-ph},
       adsurl = {https://ui.adsabs.harvard.edu/abs/2003ApJ...598..886U}
}

@ARTICLE{vandenberk2001,
       author = {{Vanden Berk}, Daniel E. and {Richards}, Gordon T. and {Bauer}, Amanda and {Strauss}, Michael A. and {Schneider}, Donald P. and {Heckman}, Timothy M. and {York}, Donald G. and {Hall}, Patrick B. and {Fan}, Xiaohui and {Knapp}, G.~R. and {Anderson}, Scott F. and {Annis}, James and {Bahcall}, Neta A. and {Bernardi}, Mariangela and {Briggs}, John W. and {Brinkmann}, J. and {Brunner}, Robert and {Burles}, Scott and {Carey}, Larry and {Castander}, Francisco J. and {Connolly}, A.~J. and {Crocker}, J.~H. and {Csabai}, Istv{\'a}n and {Doi}, Mamoru and {Finkbeiner}, Douglas and {Friedman}, Scott and {Frieman}, Joshua A. and {Fukugita}, Masataka and {Gunn}, James E. and {Hennessy}, G.~S. and {Ivezi{\'c}}, {\v{Z}}eljko and {Kent}, Stephen and {Kunszt}, Peter Z. and {Lamb}, D.~Q. and {Leger}, R. French and {Long}, Daniel C. and {Loveday}, Jon and {Lupton}, Robert H. and {Meiksin}, Avery and {Merelli}, Aronne and {Munn}, Jeffrey A. and {Newberg}, Heidi Jo and {Newcomb}, Matt and {Nichol}, R.~C. and {Owen}, Russell and {Pier}, Jeffrey R. and {Pope}, Adrian and {Rockosi}, Constance M. and {Schlegel}, David J. and {Siegmund}, Walter A. and {Smee}, Stephen and {Snir}, Yehuda and {Stoughton}, Chris and {Stubbs}, Christopher and {SubbaRao}, Mark and {Szalay}, Alexander S. and {Szokoly}, Gyula P. and {Tremonti}, Christy and {Uomoto}, Alan and {Waddell}, Patrick and {Yanny}, Brian and {Zheng}, Wei},
        title = "{Composite Quasar Spectra from the Sloan Digital Sky Survey}",
      journal = {\aj},
         year = 2001,
        month = aug,
       volume = {122},
       number = {2},
        pages = {549-564},
          doi = {10.1086/321167},
archivePrefix = {arXiv},
       eprint = {astro-ph/0105231},
 primaryClass = {astro-ph},
       adsurl = {https://ui.adsabs.harvard.edu/abs/2001AJ....122..549V}
}

@ARTICLE{Veilleux&Osterbrock1987,
       author = {{Veilleux}, Sylvain and {Osterbrock}, Donald E.},
        title = "{Spectral Classification of Emission-Line Galaxies}",
      journal = {\apjs},
         year = 1987,
        month = feb,
       volume = {63},
        pages = {295},
          doi = {10.1086/191166},
       adsurl = {https://ui.adsabs.harvard.edu/abs/1987ApJS...63..295V}
}

@ARTICLE{veron04,
       author = {{V{\'e}ron-Cetty}, M. -P. and {Joly}, M. and {V{\'e}ron}, P.},
        title = "{The unusual emission line spectrum of I Zw 1}",
      journal = {\aap},
         year = 2004,
        month = apr,
       volume = {417},
        pages = {515-525},
          doi = {10.1051/0004-6361:20035714},
archivePrefix = {arXiv},
       eprint = {astro-ph/0312654},
 primaryClass = {astro-ph},
       adsurl = {https://ui.adsabs.harvard.edu/abs/2004A&A...417..515V}
}

@ARTICLE{Volonteri2021,
       author = {{Volonteri}, Marta and {Habouzit}, M{\'e}lanie and {Colpi}, Monica},
        title = "{The origins of massive black holes}",
      journal = {Nature Reviews Physics},
         year = 2021,
        month = sep,
       volume = {3},
       number = {11},
        pages = {732-743},
          doi = {10.1038/s42254-021-00364-9},
archivePrefix = {arXiv},
       eprint = {2110.10175},
 primaryClass = {astro-ph.GA},
       adsurl = {https://ui.adsabs.harvard.edu/abs/2021NatRP...3..732V}
}

@ARTICLE{wang09,
       author = {{Wang}, Jian-Guo and {Dong}, Xiao-Bo and {Wang}, Ting-Gui and {Ho}, Luis C. and {Yuan}, Weimin and {Wang}, Huiyuan and {Zhang}, Kai and {Zhang}, Shaohua and {Zhou}, Hongyan},
        title = "{Estimating Black Hole Masses in Active Galactic Nuclei Using the Mg II {\ensuremath{\lambda}}2800 Emission Line}",
      journal = {\apj},
         year = 2009,
        month = dec,
       volume = {707},
       number = {2},
        pages = {1334-1346},
          doi = {10.1088/0004-637X/707/2/1334},
archivePrefix = {arXiv},
       eprint = {0910.2848},
 primaryClass = {astro-ph.CO},
       adsurl = {https://ui.adsabs.harvard.edu/abs/2009ApJ...707.1334W}
}

@ARTICLE{Woo18,
       author = {{Woo}, Jong-Hak and {Le}, Huynh Anh N. and {Karouzos}, Marios and {Park}, Dawoo and {Park}, Daeseong and {Malkan}, Matthew A. and {Treu}, Tommaso and {Bennert}, Vardha N.},
        title = "{Calibration and Limitations of the Mg II Line-based Black Hole Masses}",
      journal = {\apj},
         year = 2018,
        month = jun,
       volume = {859},
       number = {2},
          eid = {138},
        pages = {138},
          doi = {10.3847/1538-4357/aabf3e},
archivePrefix = {arXiv},
       eprint = {1804.02798},
 primaryClass = {astro-ph.GA},
       adsurl = {https://ui.adsabs.harvard.edu/abs/2018ApJ...859..138W}
}

@ARTICLE{xiao11,
       author = {{Xiao}, Ting and {Barth}, Aaron J. and {Greene}, Jenny E. and {Ho}, Luis C. and {Bentz}, Misty C. and {Ludwig}, Randi R. and {Jiang}, Yanfei},
        title = "{Exploring the Low-mass End of the M $_{BH}$-{\ensuremath{\sigma}}$_{*}$ Relation with Active Galaxies}",
      journal = {\apj},
         year = 2011,
        month = sep,
       volume = {739},
       number = {1},
          eid = {28},
        pages = {28},
          doi = {10.1088/0004-637X/739/1/28},
archivePrefix = {arXiv},
       eprint = {1106.6232},
 primaryClass = {astro-ph.CO},
       adsurl = {https://ui.adsabs.harvard.edu/abs/2011ApJ...739...28X}
}

@ARTICLE{zhou2006,
       author = {{Zhou}, Hongyan and {Wang}, Tinggui and {Yuan}, Weimin and {Lu}, Honglin and {Dong}, Xiaobo and {Wang}, Junxian and {Lu}, Youjun},
        title = "{A Comprehensive Study of 2000 Narrow Line Seyfert 1 Galaxies from the Sloan Digital Sky Survey. I. The Sample}",
      journal = {\apjs},
         year = 2006,
        month = sep,
       volume = {166},
       number = {1},
        pages = {128-153},
          doi = {10.1086/504869},
archivePrefix = {arXiv},
       eprint = {astro-ph/0603759},
 primaryClass = {astro-ph},
       adsurl = {https://ui.adsabs.harvard.edu/abs/2006ApJS..166..128Z}
}
\bibliographystyle{aasjournal}

\end{document}